\documentclass[sigconf,table,dvipsnames,table,10pt]{acmart}
\usepackage[utf8]{inputenc}
\usepackage[T1]{fontenc}
\usepackage{amsmath}
\usepackage{graphicx}
\usepackage{setspace}% load before caption
\usepackage[margin=8pt,skip=5pt,belowskip=0pt,font={small,stretch=0.9},labelfont=bf]{caption}% load before subcaption
\usepackage{subcaption}
\usepackage{pgf-umlsd}
\usepackage{xspace}
\usepackage[abbreviations]{foreign}  % For \ie, \eg \etc
\usepackage[binary-units,per-mode=symbol]{siunitx}
\usepackage{ifthen}
\usepackage{tcolorbox}
\usepackage{multirow}
\usepackage{tabularx,booktabs}
\usepackage{listings}
\usepackage{csquotes}
\usepackage{xcolor}
\usepackage{glossaries}
\usepackage{pifont}
\makeglossaries

\newglossaryentry{vCPU}{name=vCPU, description={virtual Central Processing Unit}}
\newglossaryentry{CPU}{name=CPU, description={Central Processing Unit}}
\newglossaryentry{VM}{name=VM, description={Virtual Machine}}
\newglossaryentry{OS}{name=OS, description={Operating System}}
\newglossaryentry{AWS}{name=AWS, description={Amazon Web Services}}
\newglossaryentry{IBM}{name=IBM, description={International Business Machines Corporation}}
\newglossaryentry{CEO}{name=CEO, description={Chief Executive Officer}}
\newglossaryentry{MB}{name=MB, description={Megabyte}}
\newglossaryentry{GB}{name=GB, description={Gigabyte}}
\newglossaryentry{MHz}{name=MHz, description={Megahertz}}
\newglossaryentry{GHz}{name=GHz, description={Gigahertz}}
\newglossaryentry{HTTP}{name=HTTP, description={Hypertext Transfer Protocol}}
\newglossaryentry{HTML}{name=HTML, description={Hypertext Markup Language}}
\newglossaryentry{CSS}{name=CSS, description={Cascading Style Sheets}}
\newglossaryentry{JSON}{name=JSON, description={JavaScript Object Notation}}
\newglossaryentry{API}{name={API}, description={Application Programming Interface}}
\newglossaryentry{RPS}{name={RPS}, description={Requests per second}}
\newglossaryentry{GUI}{name={GUI}, description={Graphical User Interface}}

\newacronym{IaaS}{IaaS}{Infrastructure as a Service}
\newacronym{PaaS}{PaaS}{Platform as a Service}
\newacronym{SaaS}{SaaS}{Software as a Service}
\newacronym{FaaS}{FaaS}{Function as a Service}
\newacronym{ACU}{ACU}{Azure Compute Unit}
\newacronym{NIST}{NIST}{National Institute of Standards and Technology}
\newacronym{NPM}{NPM}{Node Package Manager}
\newacronym{CLI}{CLI}{Command Line Interface}
\newacronym{EC2}{EC2}{Elastic Compute Cloud}
\newacronym{TSDB}{TSDB}{Time Series Database}

\definecolor{javascriptblue}{rgb}{0.10,0.10,0.66} %for variable
\definecolor{javascriptroyalblue}{rgb}{0.10,0,0.82} %for numbers
\definecolor{javascriptgreen}{rgb}{0,0.46,0} %for comments
\definecolor{javascriptbrown}{rgb}{0.6,0.27,0} %for symbols
\definecolor{javascriptpurple}{rgb}{0.67,0.05,0.56} %for keywords
\definecolor{javascriptred}{rgb}{0.77,0.09,0.08} %for strings

\newcounter{zaehler}
\theoremstyle{definition}

\theoremstyle{definition}

\newcommand{\SYS}{\textsc{FaaSdom}\xspace}
\newcommand{\sys}{\SYS} %alias

\usepackage{cleveref}
\setcopyright{none}
\renewcommand\footnotetextcopyrightpermission[1]{}

\title[\sys: A Benchmark Suite for Serverless Computing]{\sys: A Benchmark Suite\\for Serverless Computing}
\author{Pascal Maissen}
\affiliation{Department of Computer Science \\
University of Fribourg, Switzerland}
\email{pascal.maissen@unifr.ch}

\author{Pascal Felber}
\affiliation{Department of Computer Science \\
University of Neuch\^atel, Switzerland}
\email{pascal.felber@unine.ch}

\author{Peter Kropf}
\affiliation{Department of Computer Science \\
University of Neuch\^atel, Switzerland}
\email{peter.kropf@unine.ch}

\author{Valerio Schiavoni}
\affiliation{Department of Computer Science \\
University of Neuch\^atel, Switzerland}
\email{valerio.schiavoni@unine.ch}

\begin{document}
%!TEX root = main.tex
%\nocite{Pellegrini_2019}
\begin{abstract}
Serverless computing has become a major trend among cloud providers. 
With serverless computing, developers fully delegate the task of managing the servers, dynamically allocating the required resources, as well as handling availability and fault-tolerance matters to the cloud provider.
In doing so, developers can solely focus on the application logic of their software, which is then deployed and completely managed in the cloud.

Despite its increasing popularity, not much is known regarding the actual system performance achievable on the currently available serverless platforms.
Specifically, it is cumbersome to benchmark such systems in a language- or runtime-independent manner.
Instead, one must resort to a full application deployment, to later take informed decisions on the most convenient solution along several dimensions, including performance and economic costs.

\sys is a modular architecture and proof-of-concept implementation of a benchmark suite for serverless computing platforms.
It currently supports the current mainstream serverless cloud providers (\emph{i.e.}, AWS, Azure, Google, IBM), a large set of benchmark tests and a variety of implementation languages.
The suite fully automatizes the deployment, execution and clean-up of such tests, providing insights (including historical) on the performance observed by serverless applications. 
\sys also integrates a model to estimate budget costs for deployments across the supported providers.
\sys is open-source and available at \sloppy{\url{https://github.com/bschitter/benchmark-suite-serverless-computing}}.
%Among all the services the cloud offers, serverless computing is very popular and every large cloud service provider offers a serverless platform. 
%However, there has not been much effort to test, benchmark or compare these services. 
%This thesis tries to take an approach on benchmarking serverless computing with building a suite that everyone can use.

%This suite provides five different tests to benchmark on four different clouds, highly automated deployment and cleanup of these tests, a testing utility for comparing cloud providers, a heavy benchmark to load test the serverless platforms and a pricing calculator to estimate hypothetical and actual costs. 
%Everything is packaged with Docker and is easy to use.

%This benchmark suite is completely open source and the code and the documentation can be found at \url{https://github.com/bschitter/benchmark-suite-serverless-computing}.
\end{abstract}
%\textbf{Keywords:} Serverless, Serverless Computing, Benchmarking, \gls{AWS} Lambda, Microsoft Azure Functions, Google Cloud Functions, %\gls{IBM} Cloud Functions, \gls{FaaS}

\fancyhead{
     \vspace{-30pt}
     \begin{tikzpicture}
         \node[align=center] () at (0,0) {
             \begin{tcolorbox}[colback=yellow!40,
                               colframe=white,
                               width=\textwidth,
                               boxrule=0mm,
                               sharp corners]
%                \begin{center}
					 \small
                     \centering
                     CC-BY 4.0. This is the author's preprint version of the camera-ready article. A full version of this paper is published in the proceedings of 14th ACM International Conference on Distributed and Event-Based Systems (DEBS 2020).
%                \end{center}
             \end{tcolorbox}
         };
     \end{tikzpicture}
}

\maketitle
\thispagestyle{fancy}
%!TEX root = main.tex
\section{Introduction}

%This chapter explains the motivation behind the project and the importance to test and benchmark the different available platforms from cloud providers. It introduces the topics of cloud computing and particularly serverless computing and briefly explains the term benchmarking.
%\subsection{Motivation}
%In today's world cloud computing is for many companies and organizations a good and maybe the best option to set up their computing infrastructure or migrate to it. 
%Smaller or newer companies often cannot afford to invest in high-performance hardware which they also need to manage themselves. Sometimes companies don't have the knowledge to maintain hardware properly and just want their databases and web servers to work, instead of worrying about a hardware failure or a power outage. 
%Furthermore most businesses want to focus on their core business which many times means using computing resources and tools instead of managing them.\\
%This thesis is going to investigate a specific region of cloud computing; generally known as 
The serverless computing paradigm is an emerging approach for developing cloud-based applications~\cite{Baldini2017, riseofserverless, vanEyk:2017:SCG:3154847.3154848}.
\gls{IBM}~\cite{serverlessibm} defines it as \emph{``an approach to computing that offloads responsibility for common infrastructure management tasks (scaling, scheduling, patching, provisioning, \etc) to cloud providers and tools, allowing engineers to focus their time and effort on the business logic specific to their applications or process''}.
Serverless computing requires less expertise than other self-managed approaches.
Users do not manage directly the infrastructure and runtime of the system, but instead delegate its operations to the cloud provider.
Additionally, cloud providers can deploy finer-grain billing policies (\eg, on a per service-call basis) for any of the offered services~\cite{serverlessaws, serverlessazure}, generally leading to reduced costs for developers.

\begin{figure}[!t]
\begin{center}
\includegraphics[scale=0.7]{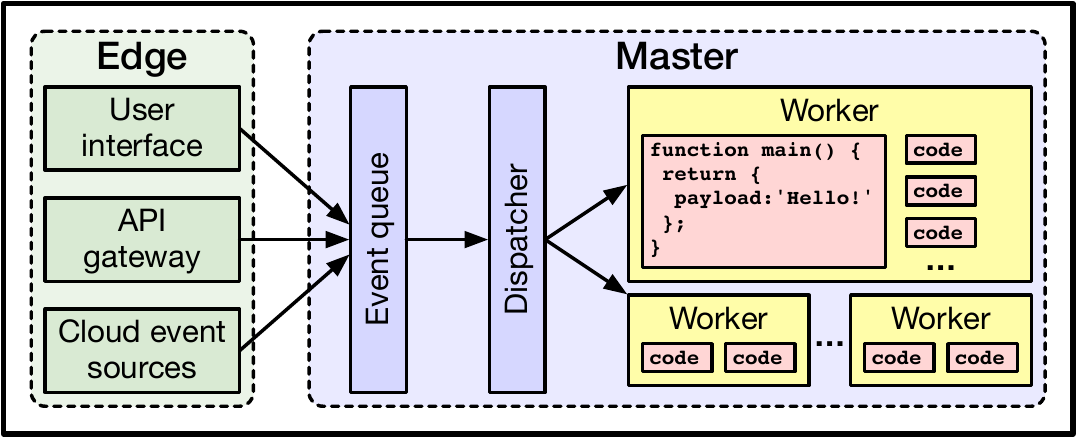}
%\captionsetup[table]{justification=centering, labelfont=bf}
\caption{Typical FaaS architecture. The main aspects to benchmark are: performance of a worker (\ie, execution speed of a function) and quality of the auto-scaling mechanism (\eg, allocation of new VMs, how to handle flash-crowds~\cite{jung2002flash}).\label{fig:faas_arch}}
\end{center}
\end{figure}
%\vs{MAISSEN: you mean time to allocate/deallocate a new VM?}
%\mais{BELONGS TO FIG 1! (Sorry I think it didn't fit into the caption) Yes mostly but not only. The main part is the time to allocate a new VM but also the question of how many regarding to the load (which e.g. requests increase drastically from 1 per second to 500 per second).} \mais{Do you know what I mean? It is more than just the time it takes, also the mechanism or decision of the cloud how many}

%Using serverless technologies requires much less expertise than non serverless self managed implementations. 
%Although those technologies might come with certain limitations or performance bottlenecks that won't fit to the user's needs.\\
%The most important key features of serverless computing are the following: 
%No server or infrastructure management of the user is required, the workload is scaled dynamically and automatically and it is usually paid per usage, e.g. only charged for the occupied storage in a service \cite{serverlessaws, serverlessazure}.\\
One can distinguish between various serverless paradigms:
\emph{(1)} FaaS (\emph{function as a service}, and focus of our work) implemented for instance by \gls{AWS} Lambda~\cite{AWSLambda},
\emph{(2)} DBaaS (\emph{database as a service}), as available through Microsoft Azure for PostgreSQL~\cite{AzureDBaaS}; and
\emph{(3)} STaaS (\emph{storage as a service}), via Google Cloud~\cite{serverlessgoogle}. 
FaaS can be considered an hybrid between \gls{PaaS} and the \gls{SaaS} service model: data and infrastructure are fully managed by the provider, while the application is handled by the user.
\autoref{fig:faas_arch} illustrates a typical FaaS infrastructure.
%\vs{Maissen: please expand description of Figure~\ref{fig:faas_arch}, mention which components require benchmarking - typically the ones that we find again in the evaluation}.
In the typical FaaS approach, developers bypass the setup, maintenance and management of a compute node (\ie, bare metal, virtual machines, or even containers). % don't need to setup or manage its operating system.
Instead, users provide the application code for specific \emph{functions} to be deployed to the cloud.
Specific events (\eg, \gls{HTTP} requests, storage or database conditions) trigger their execution, typically implementing data processing~\cite{AWSLambda, GoogleFunctions}.
The provider then handles the load, as well as availability and scaling requirements. % the application generates and other important elements like 
Despite the convenience of the approach, it is currently hard to decide on a specific FaaS provider based on criteria such as performance, workload adaptability or costs.
%Which cloud provider to choose to run an application on? Can the cloud provider handle the load? How much is it going to cost? 

This paper introduces \sys, a testing suite that tackles this problem.
In a nutshell, application developers can use the \sys suite to automatically deploy several performance tests a set of FaaS providers.
The results can then be easily compared along several dimensions.
Based on this information, deployers can evaluate the ``FaaS-domness'' of the providers and decide which one is best adapted for their applications, in terms of performance, reliability or cost.
% and the suite helps to make the decision easier.

While few efforts exist to benchmark serverless computing (often FaaS oriented) ~\cite{doi:10.1002/cpe.4792, Kuntsevich:2018:DAB:3284014.3284016, EoPSCE, 10.1007/978-3-319-75178-8_34, DBLP:conf/wosp/EykSEAI20, DBLP:conf/sac/KuhlenkampWBEW20, DBLP:conf/comsnets/SomuDBK20,DBLP:conf/asplos/GanZCSRKBHRJHPH19,DBLP:conf/cloud/JonasPVSR17,DBLP:journals/corr/abs-1912-00937,DBLP:conf/cidr/HellersteinFGSS19}, they are relatively limited in terms of supported providers, comparison metrics, diversity of benchmarks (\eg, user-defined functions, elasticity), or operating modes (\eg, stress-test vs. continuous monitoring).
% ...\vs{fill in with 1 sentence saying why those efforts are not sufficient/clear/poor/}. 
Similarly, studies on benchmarking cloud platforms lack an in-depth evaluation of serverless computing platforms~\cite{Gan:2019:OBS:3297858.3304013}. 
%\vs{
%Related work (should be compared and cited):
%\url{https://arxiv.org/pdf/1812.03651.pdf} (CIDR'19)
%\url{https://arxiv.org/pdf/1912.00937v1.pdf} (SIGMOD'20)
%\url{https://dl.acm.org/doi/10.1145/3127479.3128601} (SoCC'17)
%\url{https://serverless-benchmark.com/}
%\url{http://www.csl.cornell.edu/~delimitrou/papers/2019.asplos.microservices.pdf} (ASPLOS'19)
%}
%
In this paper, we introduce \sys, a modular and extensible benchmark suite for evaluating serverless computing.
\sys natively supports major FaaS providers (AWS, Azure, Google, IBM) but can be easily extended to benchmark additional platforms.
It includes a wide range of workloads, including user-defined functions, and implementation languages.
\sys uses several metrics to compare FaaS platforms against multiple dimensions, in terms of latency, throughput and operating costs.
It provides a Web-based interface that allows users to perform benchmarks in a fully automatic way (including deployment, execution and clean-up of the tests), and keeps track of historical data.
In that way, \sys can be used both for one-shot benchmarks and for continuous monitoring over time of the providers.
\sys is the first system to support such in-depth, comprehensive and extensible benchmarking of serverless computing.

The remainder of this paper is organised as follows.
We first introduce background concepts (\S\ref{sec:background}) and the supported frameworks (\S\ref{sec:runtime}).
We then describe the \sys architecture (\S\ref{sec:architecture}) and its different benchmarks (\S\ref{sec:tests}).
We present and discuss evaluation results (\S\ref{sec:evaluation}), before concluding with a summary of lessons learned (\S\ref{sec:lessons}) and open perspectives(\S\ref{sec:conclusion}).

\newcommand{\YES}{\textcolor{OliveGreen}{\ding{51}}}
\newcommand{\NO}{\color{BrickRed}{\ding{55}}} %\textcolor{Red4}{\ding{55}}

\renewcommand{\arraystretch}{1.2}
\begin{table*}[!t]
\scriptsize
\caption{Runtime systems supported by mainstream FaaS providers (\textsuperscript{2,3}:~generation of Azure Functions; $^\dagger$:~deprecated; $^\beta$:~beta; $^\#$:~only in \emph{App Service} or \emph{Premium} plans).}
	
\centering
% \resizebox{2\columnwidth}{!}{
\begin{tabular}{l|l|c|c|c|c|c|c|c|c|c|c|c|c|c|c|c|c|c|c|c|c} 
% \hline
 \multicolumn{2}{c|}{} & \multicolumn{4}{c|}{\textbf{Node.js}} & \multicolumn{4}{c|}{\textbf{Python}} & \multicolumn{2}{c|}{\textbf{Go}} & \multicolumn{3}{c|}{\textbf{.NET Core}} & \multicolumn{2}{c|}{\textbf{Java}} & \multicolumn{2}{c|}{\textbf{Ruby}} & \textbf{Swift} & \textbf{PHP} & \multirow{2}{*}{\textbf{Docker}} \\ \cline{3-21}
 \multicolumn{2}{c|}{} & 6.x & 8.x & 10.x & 12.x & 2.7 & 3.6 & 3.7 & 3.8 & 1.11 & 1.13 & 2.1 & 2.2 & 3.1 & 8 & 11 & 2.5 & 2.7 & 4.2 & 7.3 & \\ \hline
 \multicolumn{2}{c|}{AWS} & \cellcolor{red!25}\NO & \cellcolor{red!25}\NO & \cellcolor{green!25}\YES & \cellcolor{green!25}\YES & \cellcolor{green!25}\YES & \cellcolor{green!25}\YES & \cellcolor{green!25}\YES & \cellcolor{green!25}\YES & \cellcolor{green!25}\YES & \cellcolor{green!25}\YES & \cellcolor{green!25}\YES & \cellcolor{red!25}\NO & \cellcolor{red!25}\NO & \cellcolor{green!25}\YES & \cellcolor{green!25}\YES & \cellcolor{green!25}\YES & \cellcolor{green!25}\YES & \cellcolor{red!25}\NO & \cellcolor{red!25}\NO & \cellcolor{yellow!25}ECS \\ \hline
 \multirow{2}{*}{Azure} & Linux & \cellcolor{red!25}\NO & \cellcolor{green!25}\YES\textsuperscript{2} & \cellcolor{green!25}\YES\textsuperscript{2,3} & \cellcolor{green!25}\YES\textsuperscript{3} & \cellcolor{red!25}\NO & \cellcolor{green!25}\YES\textsuperscript{2,3} & \cellcolor{green!25}\YES\textsuperscript{2,3} & \cellcolor{green!25}\YES\textsuperscript{3} & \cellcolor{red!25}\NO & \cellcolor{red!25}\NO & \cellcolor{red!25}\NO & \cellcolor{green!25}\YES\textsuperscript{2} & \cellcolor{green!25}\YES\textsuperscript{3} & \cellcolor{green!25}\YES\textsuperscript{2,3} & \cellcolor{red!25}\NO & \cellcolor{red!25}\NO & \cellcolor{red!25}\NO & \cellcolor{red!25}\NO & \cellcolor{red!25}\NO & \cellcolor{yellow!25}\YES$^\#$ \\ \cline{2-22}
  & Windows & \cellcolor{red!25}\NO & \cellcolor{green!25}\YES\textsuperscript{2} & \cellcolor{green!25}\YES\textsuperscript{2,3} & \cellcolor{green!25}\YES\textsuperscript{3} & \cellcolor{red!25}\NO & \cellcolor{red!25}\NO & \cellcolor{red!25}\NO & \cellcolor{red!25}\NO & \cellcolor{red!25}\NO & \cellcolor{red!25}\NO & \cellcolor{red!25}\NO & \cellcolor{green!25}\YES\textsuperscript{2} & \cellcolor{green!25}\YES\textsuperscript{3} & \cellcolor{green!25}\YES\textsuperscript{2,3} & \cellcolor{red!25}\NO & \cellcolor{red!25}\NO & \cellcolor{red!25}\NO & \cellcolor{red!25}\NO & \cellcolor{red!25}\NO & \cellcolor{red!25}\NO \\ \hline
 \multicolumn{2}{c|}{Google} & \cellcolor{green!25}\YES$^\dagger$ & \cellcolor{green!25}\YES & \cellcolor{green!25}\YES$^\beta$ & \cellcolor{red!25}\NO & \cellcolor{red!25}\NO & \cellcolor{red!25}\NO & \cellcolor{green!25}\YES & \cellcolor{red!25}\NO & \cellcolor{green!25}\YES & \cellcolor{green!25}\YES$^\beta$ & \cellcolor{red!25}\NO & \cellcolor{red!25}\NO & \cellcolor{red!25}\NO & \cellcolor{red!25}\NO & \cellcolor{red!25}\NO & \cellcolor{red!25}\NO & \cellcolor{red!25}\NO & \cellcolor{red!25}\NO & \cellcolor{red!25}\NO & \cellcolor{yellow!25}Cloud Run \\ \hline
 \multicolumn{2}{c|}{IBM} & \cellcolor{red!25}\NO & \cellcolor{green!25}\YES & \cellcolor{green!25}\YES & \cellcolor{red!25}\NO & \cellcolor{green!25}\YES & \cellcolor{green!25}\YES & \cellcolor{green!25}\YES & \cellcolor{red!25}\NO & \cellcolor{green!25}\YES & \cellcolor{red!25}\NO & \cellcolor{red!25}\NO & \cellcolor{green!25}\YES & \cellcolor{red!25}\NO & \cellcolor{green!25}\YES & \cellcolor{red!25}\NO & \cellcolor{green!25}\YES & \cellcolor{red!25}\NO & \cellcolor{green!25}\YES & \cellcolor{green!25}\YES & \cellcolor{green!25}\YES \\ \hline
\end{tabular}
%}
\label{table:programming_languages}
\end{table*}
\section{Background}
\label{sec:background}

This section provides technical details about the four major mainstream serverless providers, namely Amazon Web Services (\S\ref{sec:ss:aws}), Microsoft Azure (\S\ref{sec:ss:azure}), Google Cloud (\S\ref{sec:ss:google}) and IBM Cloud (\S\ref{sec:ss:ibm}). 
%\vs{MAISSEN that the next sentence is true:} \mais{I don't understand what you want me to say is true?} 
We compare the performance of all of them in our evaluation (\S\ref{sec:evaluation}).

\subsection{Amazon Web Services Lambda}
\label{sec:ss:aws}

\gls{AWS} Lambda~\cite{AWSLambda} was released in November in 2014~\cite{AWSLambdaRelease}. 
\gls{AWS} Lambda spans 18 geographical regions, plus China~\cite{AWSRegions}. 
At the time of writing, it supports six different runtime systems and seven different programming languages~\cite{AWSLambdaLanguages}. 
Depending on the region where the function is deployed, Lambda supports up to 3,000 instances to serve the user functions~\cite{AWSLambdaScaling}. 
The memory allocated to a function instance can vary from 128\,\gls{MB} up to 3,008\,\gls{MB} in steps of 64\,\gls{MB}~\cite{AWSLambdaConfig}. 
The \gls{CPU} power increases linearly with its memory allocation.
For instance, according to the documentation, at 1,792\,\gls{MB} the function will get 1 \gls{vCPU}~\cite{AWSLambdaConfig}. %\vs{MAISSEN: why 1792?} \mais{That is just how they mention it in the docs and define 1 vCPU}

As observed in~\cite{216063}, Lambda executes functions using two different \gls{CPU}s, namely Intel Xeon E5-2666 clocked at 2.90\,\gls{GHz} and Intel Xeon E5-2680, clocked at 2.80\,\gls{GHz}.
%This information was extracted by Wang et al. from the file \texttt{/proc/cpuinfo} in the Linux operating system.
%\begin{remark} From my own experience I can tell that cloud providers generally don't like to tell their customers every detail of hardware they use or how exactly services are implemented. The first point might be true because they don't want to tell a customer what exact \gls{CPU}s he gets, because then he will certainly complain if it differs from the specification and the provider has to take responsibility. The second point should be obvious for economic and competition related reasons.
%\end{remark}
%Knowing that, one can more or less estimate the theoretical computing power that the virtual \gls{CPU} will provide. 
%\begin{remark}
%The pricing of \gls{AWS} Lambda and all the other services will be discussed in section \ref{sec:pricing} Pricing.
%\end{remark}

\subsection{Microsoft Azure Functions}
\label{sec:ss:azure}

Microsoft Azure Functions~\cite{AzureFunctions} was released publicly in November 2016~\cite{AzureFunctionsAnnouncement}. %in preview (Microsoft's term for beta) and then later in November 2016 it became generally available \cite{AzureFunctionsAnnouncement}. 
It supports five runtime systems and seven different programming languages~\cite{AzureFunctionsLanguages}.
In contrast with the other three cloud providers, Azure offers three different hosting plans~\cite{AzureFunctionsPlans}:
%Something a little different with Azure compared to the other three cloud providers is that one can select between three different hosting plans \cite{AzureFunctionsPlans}:
Azure Functions offers billing plans that adapt to the load and popularity of the deployed function (\emph{``Consumption''} plan), plans with finer-grain control over the computing instance size and pre-warming support (\emph{``Premium''} plan), and a billing plan customized on a given application needs (\emph{``App Service''} plan). 
This work only considers the consumption plan (generation 2.0~\cite{AzureFunctionsGenerations}), as it is the only one to be fully managed by the cloud provider and the most similar in terms of features to the plans from alternative providers. 
%\begin{itemize}
%\item[] \textbf{Consumption plan:} It adds and removes instances dynamically depending on the load on the function and cost only arise when functions are running. This is the most 'serverless' option among those three.
%\item[] \textbf{Premium plan:} The premium plan is similar to the consumption plan but offers more integration and control over the functions. Instance sizes can be chosen and instances can be pre-warmed. The cost is calculated with \gls{CPU} and \gls{GB} memory used per second.
%\item[] \textbf{App Service plan:} How many and on which \gls{VM}s the functions run can be decided in the App Service plan. Scaling happens manually, time based or based on metrics such as \gls{CPU} usage.
%\end{itemize}
%This thesis will only consider the consumption plan, as it is the default plan and fully managed, and therefore \textit{more} serverless than the others. Additionally, it is similar to the services of the other providers.\\
%There are currently three different generations of the service available \cite{AzureFunctionsGenerations}, this project uses generation 2.

Azure Functions can use as many as 200 instances and up to 1.5\,\gls{GB} memory~\cite{AzureFunctionsPlans}. 
The service can run either on Windows or Linux hosts, and is offered in 28 out of 46 publicly accessible regions~\cite{AzureRegions}.
Note that the consumption plan is only available in 11 regions for both Linux and Windows, hence we restrict our deployment to those in our experiments.
Computing nodes can be characterized by their \gls{ACU}, with 100 ACU roughly mapped to 1 \gls{vCPU}.
%For computing power, Azure has its own term named \gls{ACU} index, where 100 ACU roughly map to 1 \gls{vCPU}.
%The instances in the Azure functions consumption plan have an \gls{ACU} of 100 which is about the equivalent of 1 \gls{vCPU}. 
According to our investigations, we believe Azure Functions to be executed by virtual machines of type \textit{Av2}.\footnote{\url{https://docs.microsoft.com/de-ch/azure/virtual-machines/av2-series}}
%\vs{MAISSEN: which ones from \url{https://aws.amazon.com/ec2/instance-types/}? Clarify.} \mais{correct links: \url{https://docs.microsoft.com/en-us/azure/azure-functions/functions-scale} and \url{https://docs.microsoft.com/de-ch/azure/virtual-machines/av2-series}, Here they mention the ACU of 100. At the time of my writing on the overview table \url{https://docs.microsoft.com/de-ch/azure/virtual-machines/acu} I'm pretty sure Av2 were the only one that matched 100 ACU. I would probably omit that now.}, as it most closely resembles its declared \gls{ACU}~\cite{AzureFunctionsVMs}. 
These \gls{VM}s use three different \gls{CPU}s: Intel Xeon 8171M at 2.1\,\gls{GHz}, Intel Xeon E5-2673 v4 at 2.3\,\gls{GHz} and Intel Xeon E5-2673 v3 at 2.4\,\gls{GHz}~\cite{AzureFunctionsVMs}.

%\begin{remark}	
%\vs{Maybe we can move this remark in the previous section where we describe the providers ?}
%Azure has currently three different generations of its service \vs{MAISSEN: add url} \cite{AzureFunctionsGenerations} \mais{cite ok like this?}. 
%The first generation is currently in maintenance, while the two other ones remain available. 
%The available runtimes in the next section will only consider generation 2 (which is also used in this project):
%\end{remark}

\subsection{Google Cloud Functions}
\label{sec:ss:google}

%\vs{STOPPED HERE}
%On the Google Cloud Platform, the serverless service is simply called \textit{Functions} \cite{GoogleFunctions}. 
Google Functions~\cite{GoogleFunctions} was released on July in 2018~\cite{GoogleFunctionsReleases} and is available through seven out of the twenty Google regions~\cite{GoogleFunctionsLocations}.
%Compared to AWS and Azure that is two and a half years respectively one year later for the first release. 
It currently only supports three programming languages~\cite{GoogleFunctionsLanguages}, namely Node.js, Python and Go. 
While there is not a maximum number of allocated instances per single function, it only allows up to 1,000 functions to be executed concurrently~\cite{GoogleFunctionsQuotas}.
%The documentation does not mention a limit of maximum allocated instances per function. 
%However, it states that at maximum 1000 functions can be concurrently in execution \cite{GoogleFunctionsQuotas}.
\autoref{table:google_functions_cpu_ram} summarizes the options for CPU and memory combinations supported by the platform~\cite{GoogleFunctionsPricing}.
While the official documentation lacks details on the exact CPU models, a quick inspection of \texttt{/proc/cpuinfo} unveils certain details, such as \texttt{vendor\_id}, \texttt{cpu\_family} and \texttt{model}.
We only identified Intel-based processors during our experiments.
%We report on the use of Intel-based processors as well as some generation details in \texttt{/proc/cpuinfo} (i.e. cpu family 6, model 85).
%Possible options for CPU and memory allocation per instance are shown in table \ref{table:google_functions_cpu_ram} \cite{GoogleFunctionsPricing}. 
%The service is available in seven out of twenty regions~\cite{GoogleFunctionsLocations}.

\begin{table}[!t]
\small
\setlength{\tabcolsep}{1pt}
\caption{Memory/CPU configurations supported by Google Cloud Functions~\cite{GoogleFunctionsPricing}.}
%\captionsetup[table]{justification=centering, labelfont=bf}
\centering
% \resizebox{\columnwidth}{!}{
\begin{tabular}{l@{\hskip 3pt}|@{\hskip 3pt}rl@{\hskip 3pt}|@{\hskip 3pt}rl@{\hskip 3pt}|@{\hskip 3pt}rl@{\hskip 3pt}|@{\hskip 3pt}rl@{\hskip 3pt}|@{\hskip 3pt}rl} 
 \hline
	\textbf{Memory} & 128&\gls{MB}  & 256&\gls{MB}  & 512&\gls{MB}  & 1,024&\gls{MB}  & 2,048&\gls{MB} \\ 
	\textbf{CPU}    & 200&\gls{MHz} & 400&\gls{MHz} & 800&\gls{MHz} &   1.4&\gls{GHz} &   2.4&\gls{GHz} \\
	\hline
\end{tabular}
\vspace{-10pt}
% }
\label{table:google_functions_cpu_ram}
\end{table}

%Google does not mention what type of CPU they use for functions on the underlying infrastructure. 
%The CPU type could also not be extracted by Wang et al. \cite{216063} and the file \texttt{/proc/cpuinfo} does not show information on the CPU model name. 
%It shows however information for the fields \texttt{vendor\_id}, \texttt{cpu\_family} and \texttt{model}. 
%Those values reveal that Intel processors are used and can give an indication from which generation the CPU is. 
%\vs{See the example file content \ref{lst:cpuinfo} in the appendix.} 

\subsection{IBM Cloud Functions}
\label{sec:ss:ibm}

\gls{IBM} Cloud Functions \cite{IBMFunctions} is built on top of Apache OpenWhisk, an open source serverless cloud platform using Docker containers~\cite{OpenWhisk}. 
As such, in addition to Docker, it supports eight additional runtime systems~\cite{IBMRuntimes}: Node.js, Python, Swift, PHP, Go, Java, Ruby and .NET Core.
%\gls{IBM} Cloud Functions supports nine different runtimes including a Docker runtime \cite{IBMRuntimes}. 
The service is restricted to 1,000 concurrently active executions (including those enqueued for execution) per namespace~\cite{IBMLimits}. 
%However this limit can be increased for a specific business case but needs to be applied for via the ticketing system of the \gls{IBM} support \cite{IBMLimits}. 
Memory allocation can be set from 128\,\gls{MB} to 2,048\,\gls{MB} in steps of 32\,\gls{MB}. 
\gls{IBM} Cloud Functions is available in five out of the six IBM regions~\cite{IBMLocations}.\footnote{One region (Tokyo) lacks support for Cloud Foundry~\cite{IBMCloudFoundry} and as such cannot be used to deploy functions via the \gls{CLI}. We therefore did not include it in our evaluation.} 
%The documentation does not mention anything on \gls{CPU} or machines they are using. 
Our experiments revealed that some of the instances supporting the execution of the functions run on top of Intel Xeon E5-2683 v3 at 2.0\,\gls{GHz}.
%\vs{See the example file content \ref{lst:cpuinfo2} in the appendix.}

%!TEX root = main.tex

\section{Runtime Systems and Languages}
\label{sec:runtime}

This section describes the runtime systems and programming languages supported by the serverless providers described in \S\ref{sec:background}. 
For the sake of comparisons and fairness of benchmarking, we are interested in those supported by multiple cloud providers.  

\autoref{table:programming_languages} highlights that Node.js is supported by all cloud providers.
We explain this by the fact the peculiar features of the language and the event-driven nature of the runtime make it particularly fit for serverless computing, as well as by its popularity among developers and its productivity advantages.
% \vs{MAISSEN: why is that? is it because it is a cloud-native language? just very popular?} \mais{I can't tell you why. Probably because it is not compiled, package management is relatively simple and it is popular in general for webservers and APIs}. 
Python is similarly well supported, except by Azure on Windows. %\vs{MAISSEN: from \ref{table:programming_languages}, this seems not to be totally true} \mais{Why not? Every cloud supports at least 1 version so you can generally run your python code on each one. Also 3.6 and 3.7 is mostly supported}.
Microsoft's \texttt{.NET Core} lacks support from Google. 
In the remainder of this paper, we focus our comparison on the most supported runtime systems and languages, namely Node.js, Python, Go and .NET Core.
We briefly introduce them next.
%A brief summary of each runtime will be given in the following subsections.

\textbf{Node.js.}~\cite{tilkov2010node} is a JavaScript runtime built on Chrome's V8 JavaScript engine~\cite{Nodejs}. 
%As the name indicates, JavaScript is a scripting language and was first released on the 4th December 1995 by Netscape \cite{JavaScript}. 
%It was mostly used in addition to \gls{HTML} and \gls{CSS} in web browsers \cite{JavaScript}. 
The Node.js framework and the \gls{NPM} have greatly contributed to making JavaScript a popular language to implement all kinds of applications.
%\footnote{https://insights.stackoverflow.com/survey/2019\#most-popular-technologies} %\vs{MAISSEN: it'd be good to add a link, for instance to the tiobe popularity index \url{https://www.tiobe.com/tiobe-index/}} \mais{I'm not familiar with such things. But also maybe the stackoverflow survey? \url{https://insights.stackoverflow.com/survey/2019\#most-popular-technologies} see graph \emph{Programming, Scripting, and Markup Languages} and \emph{Other Frameworks, Libraries, and Tools}}. 
\autoref{table:programming_languages} shows the versions of Node.js supported by each cloud provider. 
%\begin{table}[!t]
%\centering
%\captionsetup[table]{justification=centering, labelfont=bf}
%\begin{tabular}{l|c|c|c|c} 
% \hline
% Node.js & AWS & Azure & Google & IBM \\ \hline
%6.x  & \cellcolor{red!25}no    & \cellcolor{red!25}no    & \cellcolor{yellow!25}yes  & \cellcolor{red!25}no\\ \hline
%8.x  & \cellcolor{red!25}no & \cellcolor{green!25}yes & \cellcolor{green!25}yes   & \cellcolor{green!25}yes \\ \hline
%10.x & \cellcolor{green!25}yes & \cellcolor{green!25}yes & \cellcolor{yellow!25}yes  & \cellcolor{green!25}yes \\ \hline
%12.x & \cellcolor{green!25}yes & \cellcolor{red!25}no & \cellcolor{red!25}no & %\cellcolor{red!25}no \\ \hline
%\end{tabular}
%\caption[Supported Node.js runtimes]{Supported Node.js runtimes~\cite{AWSLambdaLanguages, AzureFunctionsLanguages, GoogleFunctionsLanguages, IBMRuntimes}\\. Yellow=\textit{deprecated} and \textit{beta} for version 6.x and 10.x.}
%\label{table:nodejs}
%\end{table}
Because of its vast support from all providers, \sys will deploy all Node.js applications using version 10.x.
%In particular, \gls{AWS} uses version 10.x \cite{AWSLambdaLanguages}, Azure does not specify more than version 10 \cite{AzureFunctionsLanguages}, Google uses version 10.15.3 \cite{GoogleFunctionsRuntimes} and \gls{IBM} implements version 10.15.0 \cite{IBMRuntimes}.
%\begin{remark}
%Google has Node.js version 10 still in beta while Node.js version 8 End-of-life was on December 31, 2019 \cite{NodejsReleases}.
%\end{remark}
%Python is a universal, open-source and very popular programming language. It was released in 1991 and created by Guido van Rossum \cite{PythonIntro} and runs basically anywhere \cite{PythonAbout}. Python was designed to be very easy to learn and to have readable code \cite{PythonIntro}. It only uses indentation and whitespaces to define the scope of loops, functions and classes \cite{PythonIntro}. Because of those characteristics, developers can implement new features very fast. Cuong Do, software architect of YouTube, said: "Python is fast enough for our site and allows us to produce maintainable features in record times, with a minimum of developers." \cite{PythonQuotes}. 

\textbf{Python} is supported in multiple versions. % shows the support offered by the various cloud providers to the Python runtime. % which of the four cloud provider supports which Python versions.
At the time of this writing, all cloud providers support version 3.7, hence we rely on this version for our benchmarking results.

%\begin{table}[!t]
%\centering
%\captionsetup[table]{justification=centering, labelfont=bf}
%\begin{tabular}{l|c|c|c|c} 
% \hline
% Python & AWS & Azure & Google & IBM \\ \hline
%2.7  & \cellcolor{green!25}yes    & \cellcolor{red!25}no    & \cellcolor{red!25}no  & \cellcolor{green!25}yes\\ \hline
%3.6  & \cellcolor{green!25}yes & \cellcolor{green!25}yes & \cellcolor{red!25}no   & \cellcolor{green!25}yes \\ \hline
%3.7 & \cellcolor{green!25}yes & \cellcolor{green!25}yes & \cellcolor{green!25}yes  & \cellcolor{green!25}yes \\ \hline
%3.8 & \cellcolor{green!25}yes & \cellcolor{red!25}no & \cellcolor{red!25}no  & \cellcolor{red!25}no  \\ \hline
%\end{tabular}
%\caption[Supported Python runtimes]{Supported Python runtimes~\cite{AWSLambdaLanguages, AzureFunctionsLanguages, GoogleFunctionsLanguages, IBMRuntimes}}
%\label{table:python}
%\end{table}

\textbf{Go}~\cite{GoProject} is supported by the FaaS providers for two recent releases of the language.
We use version 1.11 in our evaluation.

%It is a relatively new language and was first released in March 2012 \cite{GoProject}. 
%It is compiled and therefore more efficient than interpreted languages and it has good concurrency mechanisms to benefit from today's multi-core architecture \cite{GoDoc}. 
%Go also has a garbage collector in contrast to C \cite{GoDoc}. 
%On the Go website the following statement can be found: "It's a fast, statically typed, compiled language that feels like a dynamically typed, interpreted language." \cite{GoDoc}. 
%Today, Go is a popular language given that is is easy to learn and write like Python but also very efficient like C. 
%Table~\ref{table:programming_languages} shows the supported Go versions.
%\begin{table}[!t]
%\centering
%\captionsetup[table]{justification=centering, labelfont=bf}
%\begin{tabular}{l|c|c|c|c} 
% \hline
% Go & AWS & Azure & Google & IBM \\ \hline
%1.11  & \cellcolor{green!25}yes    & \cellcolor{red!25}no    & \cellcolor{green!25}yes  & \cellcolor{green!25}yes\\ \hline
%\end{tabular}
%\caption[Supported Go runtimes]{Supported Go runtimes~\cite{AWSLambdaLanguages, AzureFunctionsLanguages, GoogleFunctionsLanguages, IBMRuntimes}. AWS supports all versions of Go 1.x, depending on the version the binary was compiled with and deployed to Lambda.}
%\label{table:go}
%\end{table}
%\vs{MAISSEN: move into the evaluation setup section:} \mais{Sorry I don't know what you want me to move} This thesis will use Go 1.11 for its benchmark functions.

\textbf{.NET Core} is not uniformly supported by all the cloud providers. 
%.NET Core is a free and open-source software framework developed by Microsoft and its community \cite{.NETCore}. The first version was released in June 2016 \cite{.NETCoreReleases}. It supports as languages C\texttt{\#}, F\texttt{\#} and Visual Basic \cite{.NETAbout} and is currently on version 3.1 which was released in December 2019 \cite{.NETCoreBlog}. 
%Microsoft has declared its love for Linux when Satya Nadella in the end of 2014 at a Microsoft Cloud Briefing in San Francisco presented a slide which stated "Microsoft  Linux". 
%\cite{MicrosoftCloudBlog}. 
%Since then the company has embraced Linux, open-source software and cross-platform support. 
%Table~\ref{table:dotnet} shows the supported .NET Core versions.
Hence, \sys uses 2.1 on \gls{AWS} and 2.2 on Azure and \gls{IBM}.
All \texttt{.NET} Core functions are implemented in the C\texttt{\#} dialect of the \texttt{.NET} framework.

%\begin{table}[!t]
%\centering
%\captionsetup[table]{justification=centering, labelfont=bf}
%\begin{tabular}{l|c|c|c|c} 
% \hline
% .NET Core & AWS & Azure & Google & IBM \\ \hline
%2.1  & \cellcolor{green!25}yes    & \cellcolor{red!25}no    & \cellcolor{red!25}no  & \cellcolor{red!25}no\\ \hline
%2.2  & \cellcolor{red!25}no    & \cellcolor{green!25}yes    & \cellcolor{red!25}no  & \cellcolor{green!25}yes\\ \hline
%3.1  & \cellcolor{red!25}no    & \cellcolor{red!25}no    & \cellcolor{red!25}no  & \cellcolor{red!25}no\\ \hline
%\end{tabular}
%\caption[Supported .NET Core runtimes]{Supported .NET Core runtimes~\cite{AWSLambdaLanguages, AzureFunctionsLanguages, GoogleFunctionsLanguages, IBMRuntimes}}
%\label{table:dotnet}
%\end{table}

%!TEX root = main.tex
\section{Architecture}
\label{sec:architecture}

\begin{figure}[!t]
\begin{center}
\includegraphics[scale=0.7]{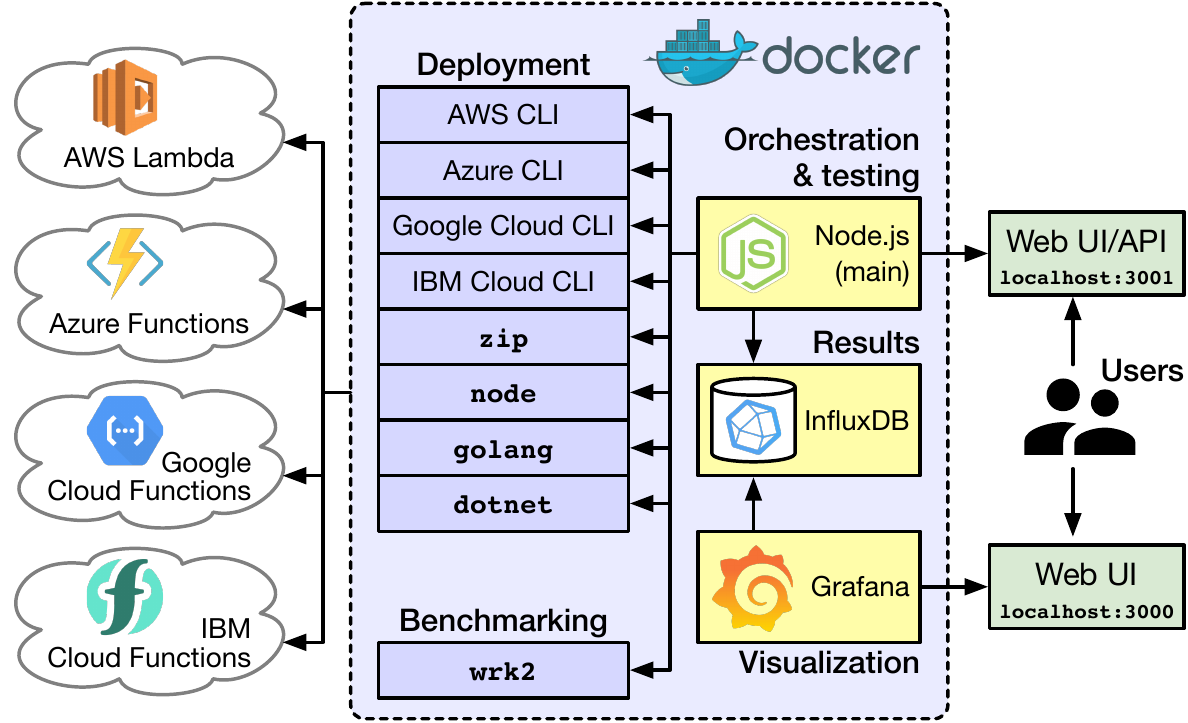}
\caption{\sys architecture.}
\label{fig:architecture}
\end{center}
\end{figure}

This section describes the architecture of the \sys prototype, depicted in \autoref{fig:architecture}, as well as providing some additional implementation details (\S\ref{ssec:impl}).
%In this chapter, the implementation of the main application, whose main role is to  orchestrate the execution of the various benchmarks across the cloud providers. 
%We provdie an overview of the varius compoentns and functionalities the components and their functions will be briefly explained and later every main task the user can do will be explained in detail.
%\subsection{Overview}
The architecture includes the supported clouds and the corresponding serverless services (left), the involved Docker images (middle) and their interface with the system (right).
We detail each component next. 

\textbf{Main application.}
The \sys core component is implemented in JavaScript and leverages the Node.js framework. 
It manages all user input and executes the actions or delegates them to other components. 
This application is packaged and executed as Docker containers. 
%Only the Linux packages \texttt{docker-ce} and \texttt{docker-compose} are needed to execute this program. 
%Figure \ref{fig:architecture} depicts all components and gives an overview of the \sys suite.
It manages the following main tasks:
\emph{(1)}~deployment to the clouds;
\emph{(2)}~execution of tests benchmarks; and
\emph{(3)}~computing price estimations. 
Users access it through a Web-based \gls{GUI} or via a REST \gls{API}. 
Once started, the set of configured tests are deployed for execution (some examples are given in \S\ref{sec:tests}).
%After the tests are deployed, they can be tested and benchmarked. 

\textbf{Time series DB.}
\sys uses a time series database (\gls{TSDB}) to store all the results from tests.
These results are subsequently used by graphical interfaces and pricing calculation.
Our prototype uses the InfluxDB~\cite{influx} \gls{TSDB}.

\textbf{UI.}
The \sys architecture provides an API to easily integrate visualization tools.
Our prototype integrates with Grafana~\cite{grafana}, an open source tool to display, plot and monitor data stored in a database. 
The results gathered by the tests and stored in InfluxDB are then displayed in Grafana.

\textbf{CLIs.}
With the exception of AWS, all cloud providers offer a Docker image for their \gls{CLI}. 
Resources can be deployed, deleted and managed completely by the \gls{CLI}.

\textbf{Runtime and languages (Node.js, Go, .NET).}
In addition to the source code of the function(s) to execute, a pre-built and packaged zip file is commonly required to successfully complete the deployment. 
The \sys architecture allows developers to ship runtime images, necessary for instance to install packages and build/run the corresponding code.

%\vs{to remove:
%\textbf{zip:} this image zips the files that need to be deployed to the clouds. 
%Therefore there is no need to have zip installed on the machine the benchmark suite runs.
%}

\textbf{Workload Injectors}.
The \sys architecture provides hook points for plugin workload injectors.
The evaluation results shown in \S\ref{sec:evaluation} uses \texttt{wrk2}~\cite{wrk2}, a stress tool for REST services, to issue requests toward services and gather throughput/latency results.

\begin{figure}[!t]
\begin{center}
\fboxrule=0.25pt\fbox{\includegraphics[scale=0.25]{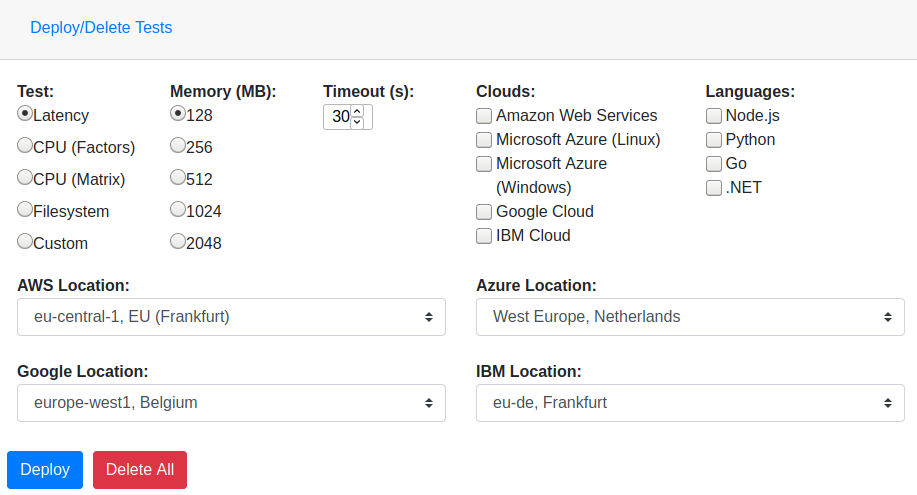}}
\caption{\sys Web-based GUI.}
\label{fig:ui}
\end{center}
\end{figure}

\begin{figure}[!b]
\begin{center}
\includegraphics[scale=0.7]{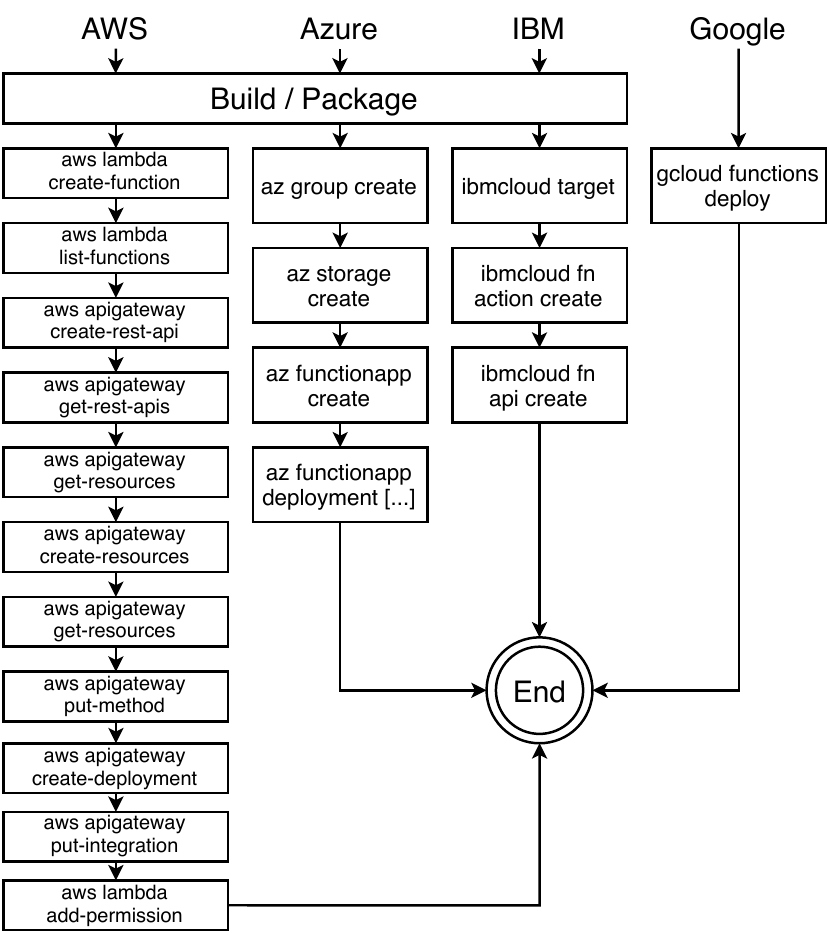}
\caption{Deployment steps for the different cloud providers supported by \sys.}
\label{fig:flowchart}
\end{center}
\end{figure}

\subsection{Deployment}

%This section will describe the deployment process of the tests.
Once the main application has started, the client can deploy the desired tests, for instance via the Web-based interface shown in \autoref{fig:ui}.
Functions can be parametrized with memory allocated for each instance, timeouts, list of providers and runtime versions to use for the benchmarks, as well as geographical regions (as supported by the cloud provider) where to deploy the tests. 
%\begin{itemize}
%    \item \textbf{Test:} a unique identifier for the test to execute to deploy;
%    \item \textbf{Memory:} The amount of memory to allocate per function.\footnote{Azure assignmes memory dynamically, up to 1536 \gls{MB} per function}
%    \item \textbf{Timeout:} The time limit after which a running function will time out. %\\ \textbf{Remark:} Not applicable for Azure since it handles timeout configuration in a configuration file.
%    \item \textbf{Clouds:} The cloud providers the tests will be deployed on, multiple choices possible.
%    \item \textbf{Languages:} Runtimes respectively languages to deploy the function in, multiple choices possible.
%    \item \textbf{Locations:} The region where the function will be deployed to.
%\end{itemize}
Once deployed, the interface continuously reports on the progress and potential issues (\eg, timeouts, access problems, \etc). 
%Next the \textit{Deploy} button can be pressed and the application will initiate the deployment. On the right hand side of the web interface there will be information about the progress. 
The deployment process is parallelized, including the creation of the required cloud resources, the build and packaging of the functions, and their upload over the cloud.
The deployment flow is slightly different for each provider, with some providers requiring significantly more operations than others.
\autoref{fig:flowchart} illustrates the various steps involved in the deployment for the providers supported by \sys.

Upon cleanup, all deployed functions are permanently deleted, as well as all the configurations and resource groups created at deployment time.

\subsection{Testing}

\begin{figure}[!t]
\begin{center}
\includegraphics[width=0.48\textwidth]{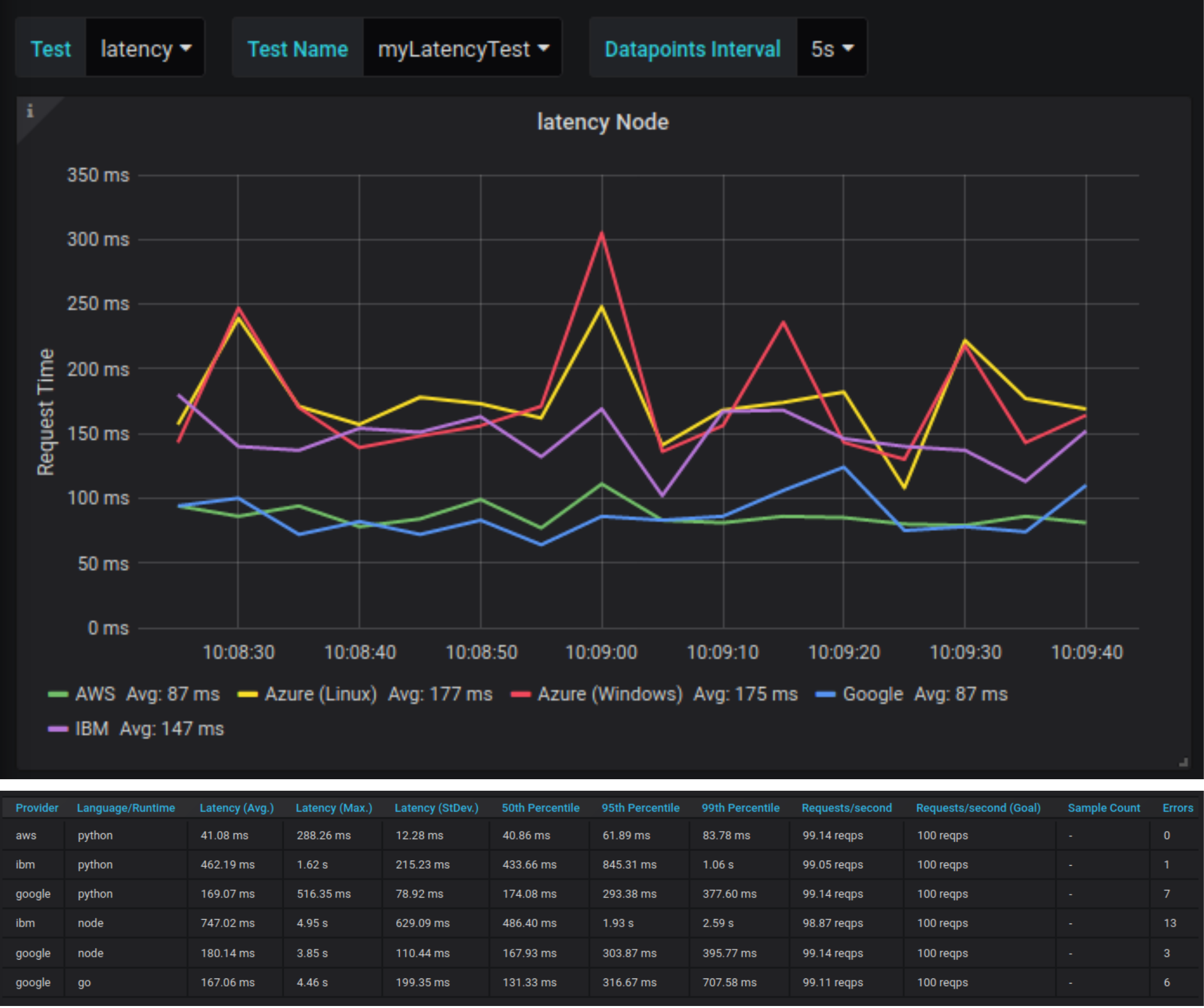}
\caption{Visualization of performance results in Grafana.}
\label{fig:grafana}
\end{center}
\end{figure}

%In this section it is briefly explained how to execute a test and see its results. 
When testing a serverless execution, \sys will send every five seconds a request to the functions previously deployed, with the purpose to gather baseline results under low, constant load. %As options, a name for the test and the functions parameter can be set. 
\autoref{fig:grafana} illustrates a screenshot of Grafana displaying the results of a latency test in Node.js. 
%At the top, there are three drop down menus to choose the test type, the test name (given at the start of the test) and the data points interval.

%The plots can be grouped by cloud provider or by runtime. 
%There is a predefined Grafana dashboard for both options, the user can switch depending on his interests. 

%An example result of a general test can be seen in section \ref{sec:general_test}.

%\begin{figure*}[!t]
%\begin{center}
%\includegraphics[width=1\textwidth]{bilder/benchmark_table.png}
%\caption[Benchmark results in Grafana]{Benchmark results in Grafana.}
%\label{fig:benchmark_table}
%\end{center}
%\end{figure*}

\subsection{Benchmarking}

For evaluating the performance of a deployed function, \sys's benchmarking component injects specific workloads toward the function exposed by a given serverless provider.
Our design is modular and currently supports \sloppy{\texttt{wrk2}~\cite{wrk2}}, a constant throughput, correct latency HTTP benchmarking tool. 
Additional tools are easy to integrate by providing a Docker container and a simple REST interface to exchange parameters and results. 
The \sys architecture allows users to directly visualize and analyze these results, using Grafana (\autoref{fig:grafana}) or other tools that can fetch the result data from the \gls{TSDB}.
%mainly relies on the wrk2 image. The parameters can be set in the web interface but they are basically just forwarded to wrk2. For the benchmark itself the user can choose the following parameters: requests per second, duration of the benchmark and the desired test to run e.g. the \gls{CPU} factors test. 
%Afterwards the load test will start and benchmark the chosen function on each cloud and runtime it was deployed to. 
%This process has to run sequentially otherwise the host of the benchmark suite could potentially not handle the load. 
%After the test has completed results will be parsed and inserted into the database and can be viewed as a table in Grafana. Figure \ref{fig:benchmark_table} shows an excerpt of a result.

% \subsection{Cleanup}

% Upon cleanup, all deployed functions are permanently deleted, as well as all the configurations and resource groups created at deployment time.
% and \gls{API} gateways on \gls{AWS} and IBM, all resource groups with a name containing latency, factors, matrix, filesystem and custom and on Google all functions in the configured project. 
%It should therefore be used very carefully and ideally with separate accounts only for this purpose.

%A more detailed test example with increasing load is discussed in section \ref{sec:loadtest}.

\subsection{Billing Costs Calculator}
\label{ssec:billingcalc}

\sys provides a pricing calculator component, which can be used to evaluate beforehand the cost of executing a certain workload on the supported cloud providers. 
%It does so either by setting the corresponding parameters manually or by exploiting results from previous executions.
%This component can be used to predict the cost of executing a given benchmark on a specific cloud provider.
Developers provide the planned workload (\eg number of function invocations, execution time per call, size of the returned data and allocated memory, \etc).
The \sys prototype produces an overview of the billing costs across the various serverless providers.
In future work, we envision this module to be able to forecast the billing costs even for full-fledged serverless applications, by applying machine-learning techniques to the system traces produced from sample executions serving real-world workloads.
%The prices for all clouds will be calculated and displayed in a table. 
%The functionality is basically the same as the pricing calculators provided by the cloud providers, except here all is in one place and directly comparable.
%Furthermore, the calculator takes the performed tests into account. 
%One can select a previously run test, select a runtime and then one only needs to provide the estimated number of invocations per month. 
%The calculation will happen by taking the execution time from the test results which of course can vary quite a lot between cloud providers and runtimes. 
%This method allows a much better approach of estimated cost. 
%In section \ref{sec:pricing} the same example will be explained and calculated for both hypothetical and with actual test results as input.

\subsection{Container Implementation}
\label{ssec:impl}

The implementation of \sys extensively relies on Docker containers.
Specifically, the main container invokes other containers using a technique called \emph{Docker-in-Docker}.\footnote{\url{https://www.docker.com/blog/docker-can-now-run-within-docker/}}
We do so by granting the main container access to \texttt{/var/run/docker.sock} and mounting it as a volume.
During our evaluation, we did not observe any particular performance degradations using this approach.

The implementation of \sys is freely available to the open-source community at \url{https://github.com/bschitter/benchmark-suite-serverless-computing}.
Its core components consist of about 2,000 lines of Node.js code.
%!TEX root = main.tex
\begin{figure*}[!t]
\centering
\includegraphics[width=1.0\textwidth]{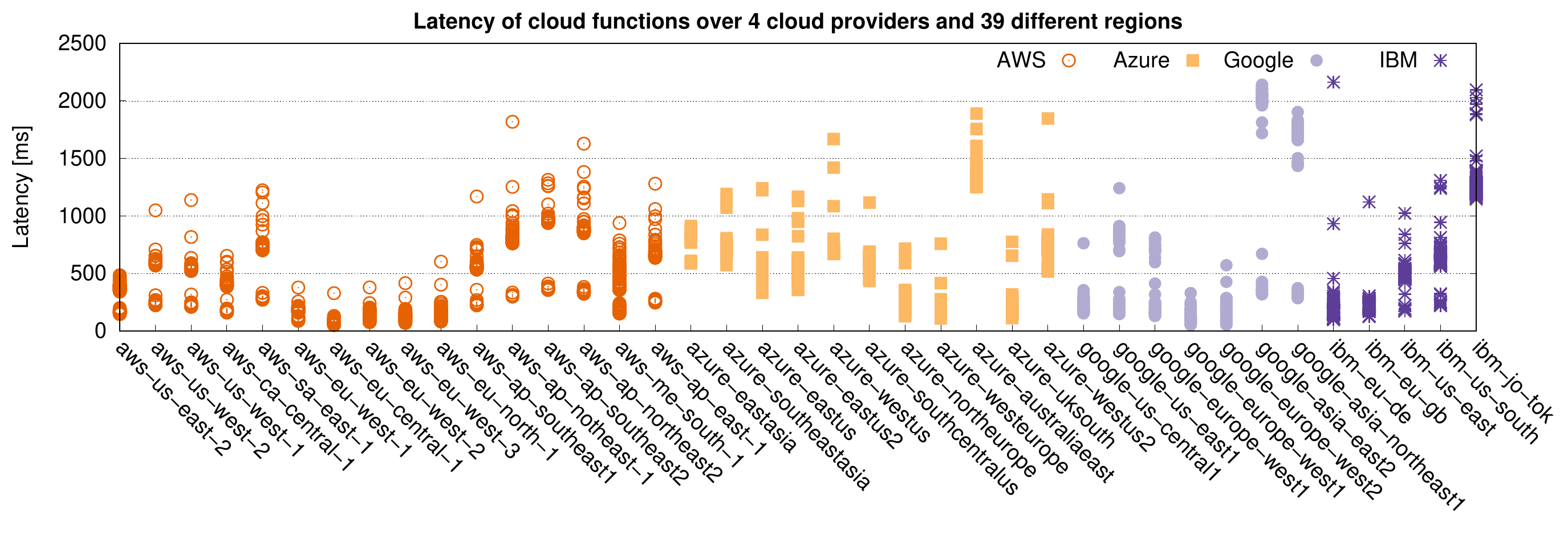}
\caption{Latency test scatter plot. Since \texttt{azure-west-us} had issues with Node.js, we resort to \texttt{.NET} for that zone.}
\label{fig:latency_plot}
\end{figure*}
\section{The \sys Benchmark Suite}
\label{sec:tests}

The \sys benchmark suite consists of a collection of tests targeting different system aspects.
Each test is implemented in several programming languages in order to support the variety of the available runtime systems.
While there are some subtle differences across the code variants to adapt to the specificities of individual cloud providers, the code executed for each programming language is largely the same for all.
All functions are implemented using an \gls{HTTP} trigger, as supported by all the serverless providers.
Currently, \sys includes the followings benchmarks:

\textbf{\texttt{faas-fact}} and \textbf{\texttt{faas-matrix-mult}} are two CPU-bound benchmarks to respectively factorize an integer and multiply large integer matrices.

\textbf{\texttt{faas-netlatency}} is a network-bound test that immediately returns upon invocation, with a small \gls{JSON} payload (HTTP body of 79 bytes and HTTP complete response including headers of $\sim$500 bytes). This test is used to verify the roundtrip times for geographically distributed deployments

\textbf{\texttt{faas-diskio}} is an IO-bound benchmark to evaluate the performance of a disk. 
Perhaps surprisingly, each serverless cloud provides a temporary file-system that functions can use to read and write intermediate results. 
Other functions executed by the same instance share this file-system to access its content.

\textbf{\texttt{faas-custom}} is provided by \sys to allows developers to test custom functions. 
The suite provides templates for the currently supported cloud providers and the supported implementation languages. 
In the long term, we expect the number of custom functions to greatly outnumber those provided by the suite itself.

\section{Evaluation}
\label{sec:evaluation}

We present in this section the results of our evaluation using \sys of the different FaaS providers supported by our current implementation.
% Rather than benchmarking the internal components of \sys, this section presents the typical evaluation results that be easily produced using \sys.
We carried out these tests over a period of three weeks (Jan. 13--Feb. 4, 2020).
These experiments specifically aim at answering the following questions: 
\emph{(1)} what is the most effective programming language to use for serverless applications;
\emph{(2)} what is the most convenient provider; and
\emph{(3)} which provider yields the most predictable results.

%chapter results of performed tests will be presented and explained, an example for a pricing calculation will be given, the services among the cloud providers will be compared and some general advantages and disadvantages about serverless computing will be specified derived from these results.
%
%In this thesis, four different tests have been performed. Each one will be discussed in detail and results will be presented in the following subsections.
\begin{figure}[!t]
\centering
\includegraphics[scale=0.7]{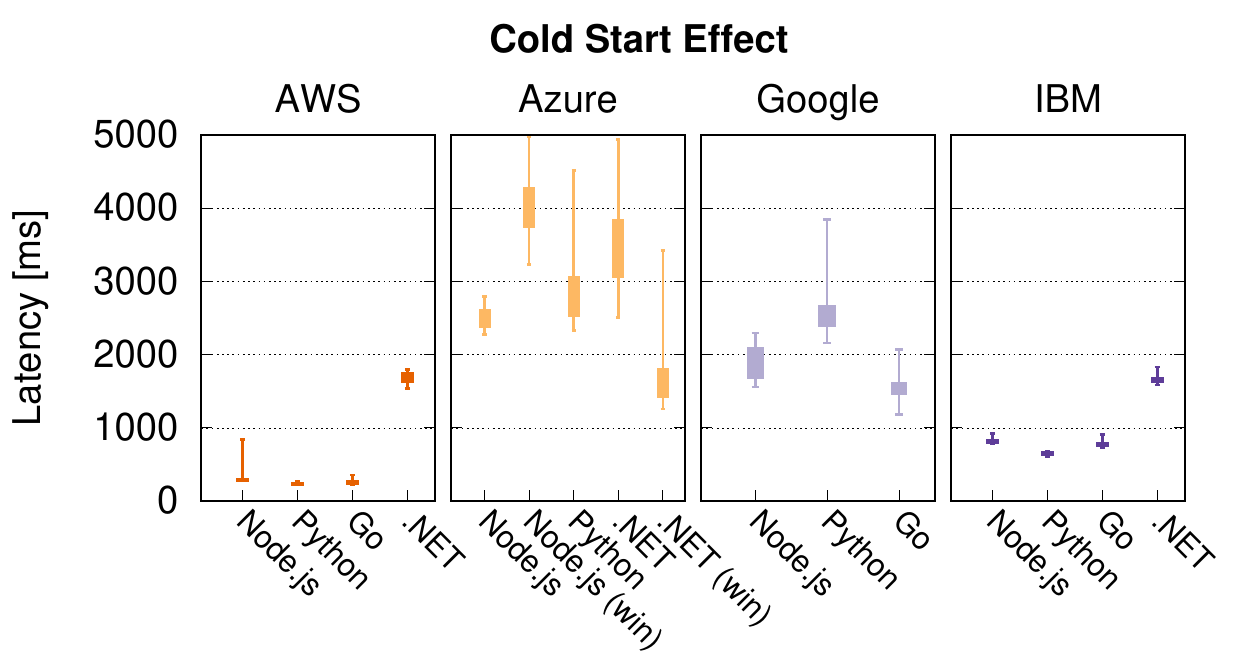}
\caption{Cold start latency.}
\label{fig:coldstart_plot}
\end{figure}
\subsection{Call Latency}
%\vs{clarify these are valid for the region from where we computed those}
We begin by measuring the latency (round-trip) for all cloud providers and corresponding regions, using the \texttt{faas-netlatency} benchmark.
%This test differs from a normal ping, that effectively a cloud function is executed instead of just returning a message on a more abstract level. 
Unless specified otherwise, we deploy Node.js using 128\,\gls{MB}.
We note that AWS functions run under a custom kernel 4.14.138-99.102.amzn2.x86\_64;\footnote{\url{https://docs.aws.amazon.com/lambda/latest/dg/lambda-runtimes.html}} 
whole host configurations are not disclosed for Azure, Google or IBM deployments.
%\mais{Azure, Google and IBM do not mention anything about the OS. Unfortunately that did not come to my mind when testing it. I can try to find things out by executing commands like 'uname -a'. Please let me know if I shall do that.}.   
%The test was performed in Node.js with 128\gls{MB} of memory (Azure 1.5\gls{GB}) with the latency test described in section \ref{subsec:latency}. 
A request is sent every 5 seconds to each cloud and region, up to 100 samples per function.
\autoref{fig:latency_plot} reports our results across 39 different configurations. 
%This test was carried out from Bern, Switzerland.
%\begin{remarks}
%\text{ }
%\begin{itemize}
%    \item For Azure, this test was only done with Linux as underlying OS. The OS should not matter in regards to latency.
%    \item The \textit{westus} region of Azure had problems deploying it in Node.js and therefore .NET was used. Following error was produced: \texttt{The scale operation is not allowed for this subscription in this region. Try selecting different region or scale option.}\\
%    Also trying to create the function in the Azure portal failed and an error was indicated but the error message could not be viewed.
%    \item On Azure region \textit{australiaeast} the function could not be deployed, although no error was risen. Therefore it was deployed on Windows instead of Linux.
%\end{itemize}
%\end{remarks}
%The graphic \ref{fig:latency_plot} shows a scatter plot with the results of the test.
%\vs{MAISSEN: here we need a short discussion on the results. Something like the following, can you complete ? I'll check afterwards}. %\mais{OK?}
%We observe the following.
The average latency over all clouds and regions is 538\,ms.
We achieve the fasted result on AWS Lambda (\texttt{eu-central-1}) at 80\,ms, and the slowest toward Google (\texttt{asia-east2}) at 1,770\,ms. 
Overall, the best performing cloud provider is Amazon, with slightly lower latency for the same geographical locations of other providers. 
Note that this benchmark was performed from Bern, Switzerland, hence different results might be achieved from different locations. 

%For more detailed results see table \ref{tab:latency}. 
%The raw result file, R script and plot image are also on \href{https://github.com/Bschitter/benchmark-suite-serverless-computing/tree/master/results/1-latency}{GitHub}.

\begin{figure}[!t]
\centering
\includegraphics[scale=0.6]{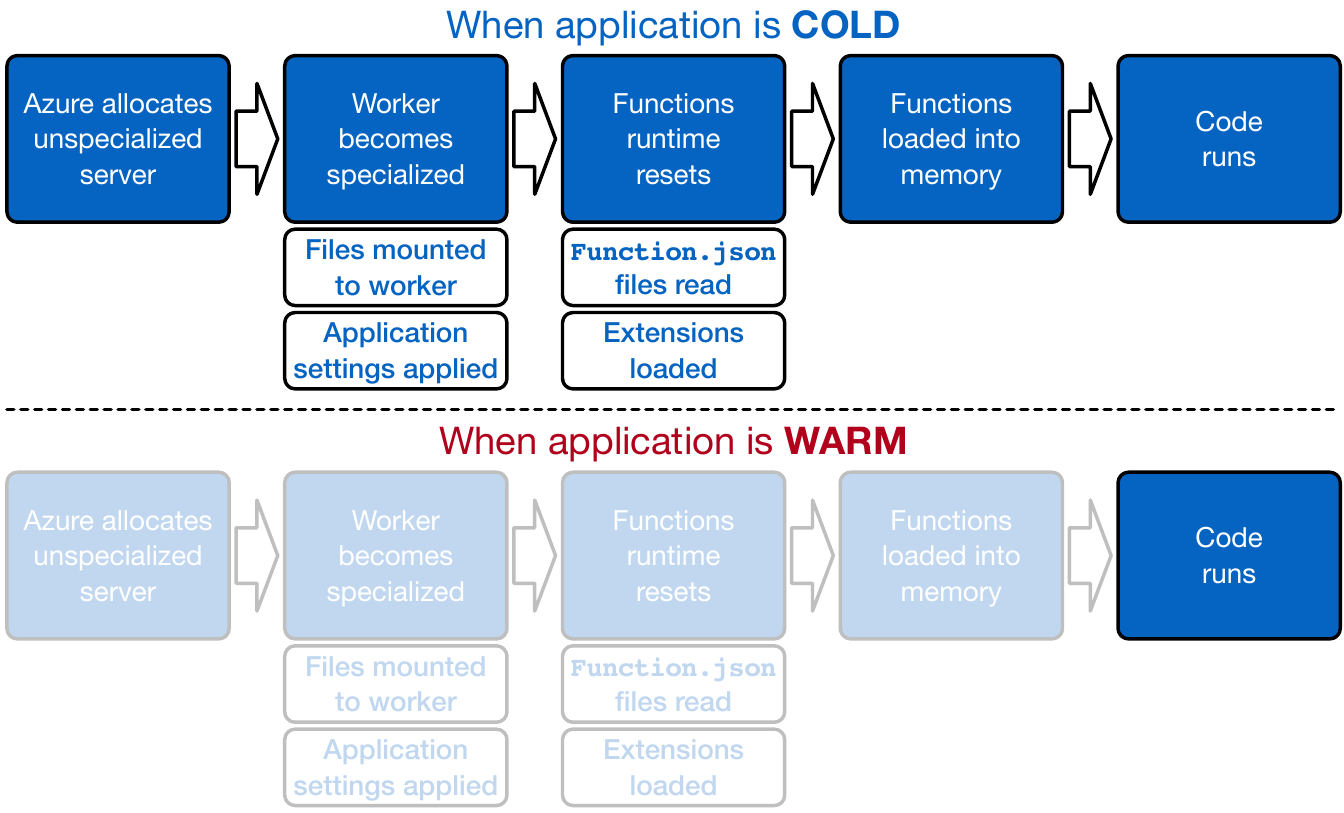}
\caption{Workflow for cold and warm start on Azure~\cite{AzureColdStart}.}
\label{fig:azure_coldstart}
\end{figure}

\begin{figure}[!b]
\centering
\includegraphics[scale=0.7]{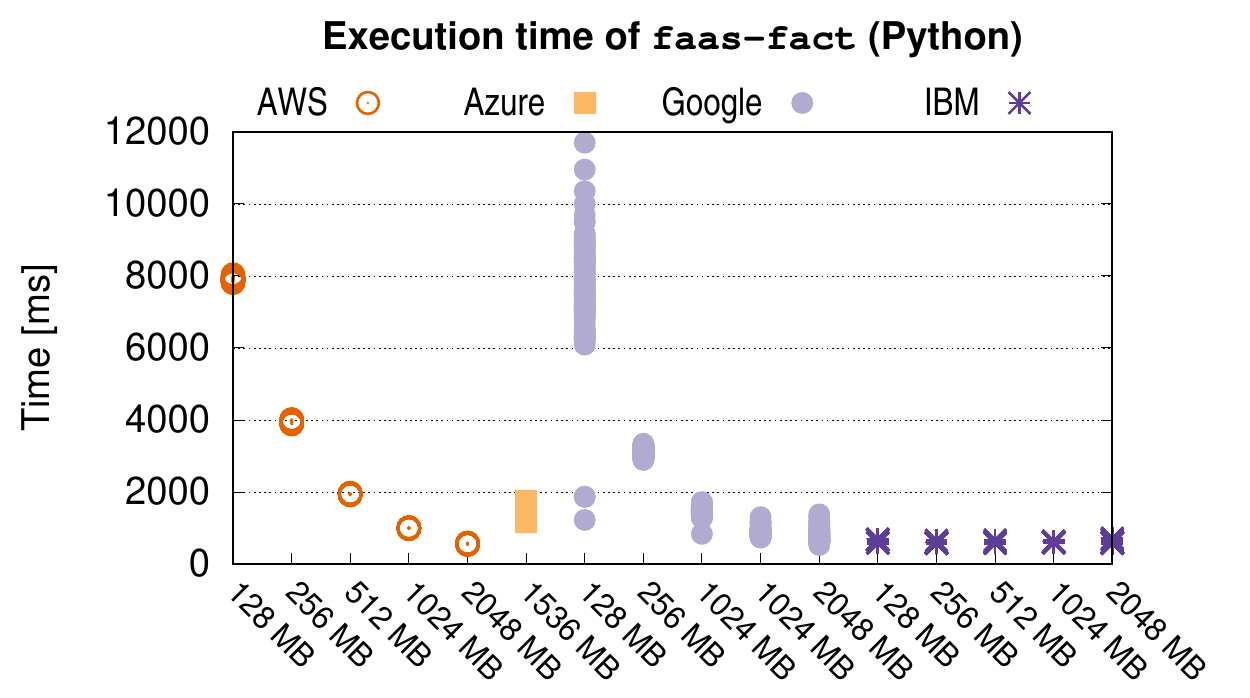}
\caption{Execution time of \texttt{faas-fact} in Python.}
\label{fig:general_python_plot}
\end{figure}

\begin{figure}[!b]
\centering
\includegraphics[scale=0.7]{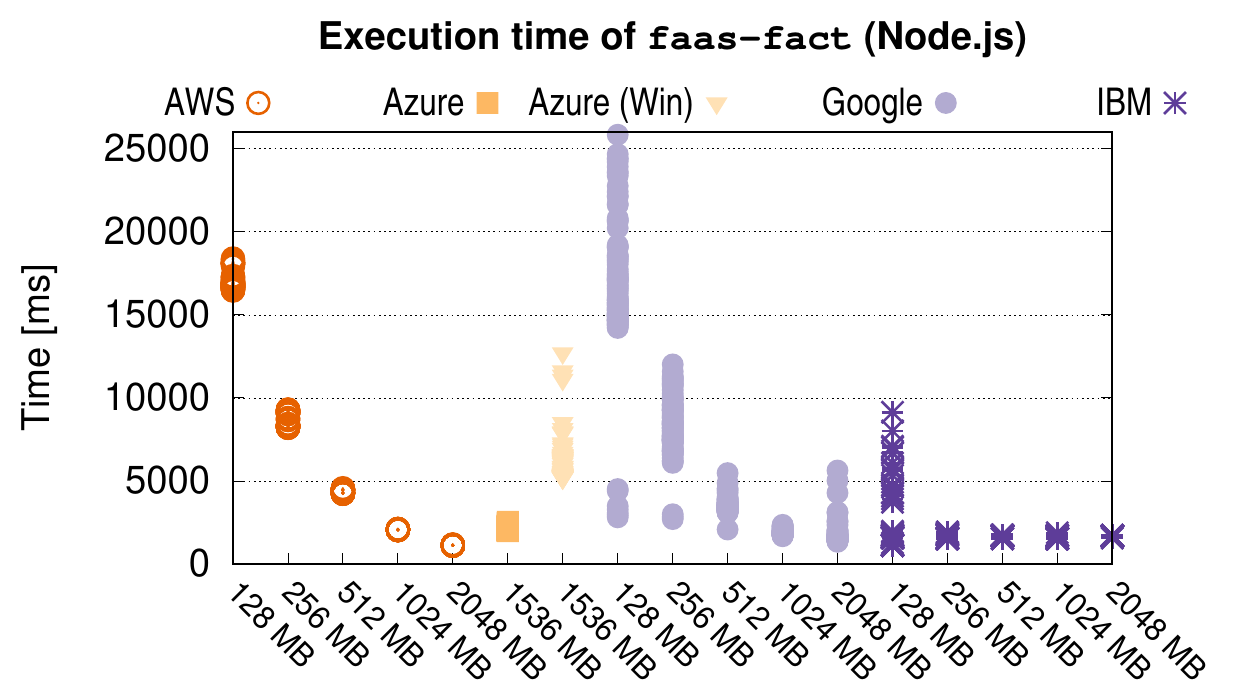}
\caption{Execution time of \texttt{faas-fact} in Node.js.}
\label{fig:general_node_plot}
\end{figure}

\subsection{Cold Start}
\label{sec:coldstart}
%\vs{clarify experimental methodology, provide further details if possible}
%This test measured the cold start latency. 
A cold start happens when a function takes longer than usual to execute. 
Most of the time, this occurs when the invocation is scheduled right after the completion of its deployment, or when the function has not been used in a while, \eg, after 10 minutes without invocations.
%However, we observed longer timings, for instance on Google, with up to 10 hours before de-allocating the instance. \mais{I don't understand the previous sentence.}
Specifically, there are several steps to perform before a function can execute: allocate a server, set up a worker with the associated code, packages and extensions, load the function in memory, and finally run it.
These steps contribute clearly to the cold start effect that we observe in our results.
When the function is \emph{warm}, it is ready in memory and can immediately serve requests.
A function can also be de-allocated (\ie, garbage-collected) when a new instance needs to be provisioned for scaling purposes.

\begin{figure}[!b]
\centering
\includegraphics[scale=0.7]{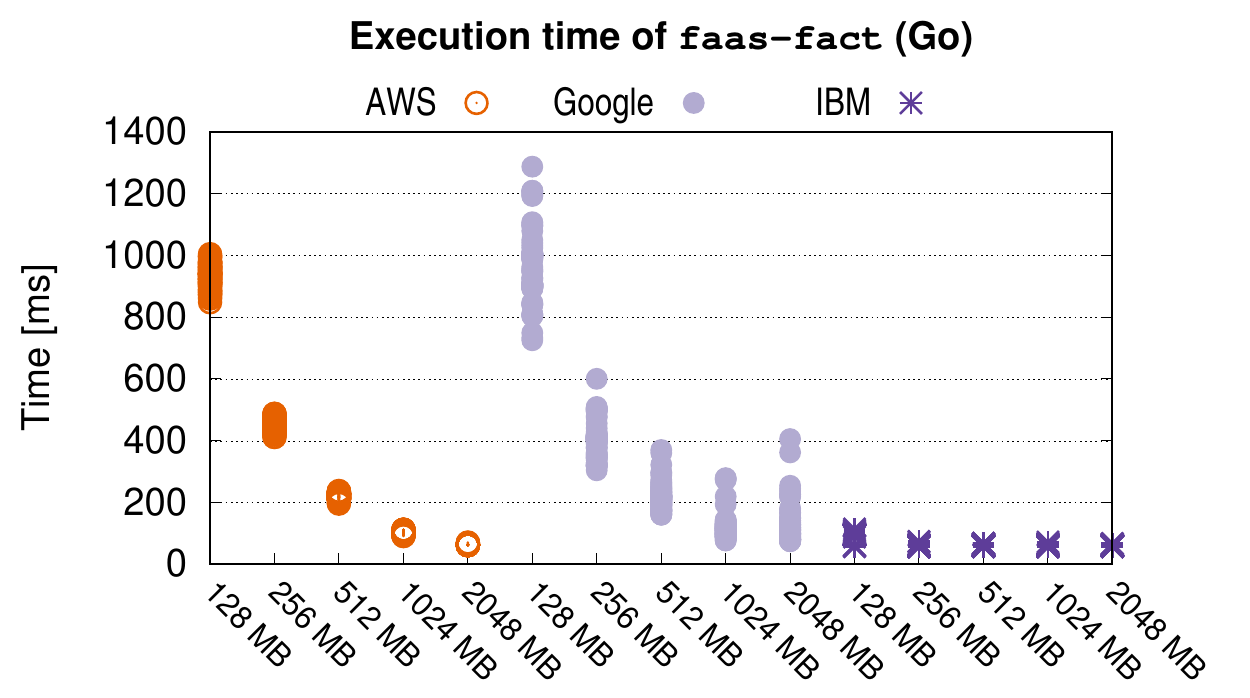}
\caption{Execution time of \texttt{faas-fact} in Go.}
\label{fig:general_go_plot}
\end{figure}

\begin{figure}[!b]
\centering
\includegraphics[scale=0.7]{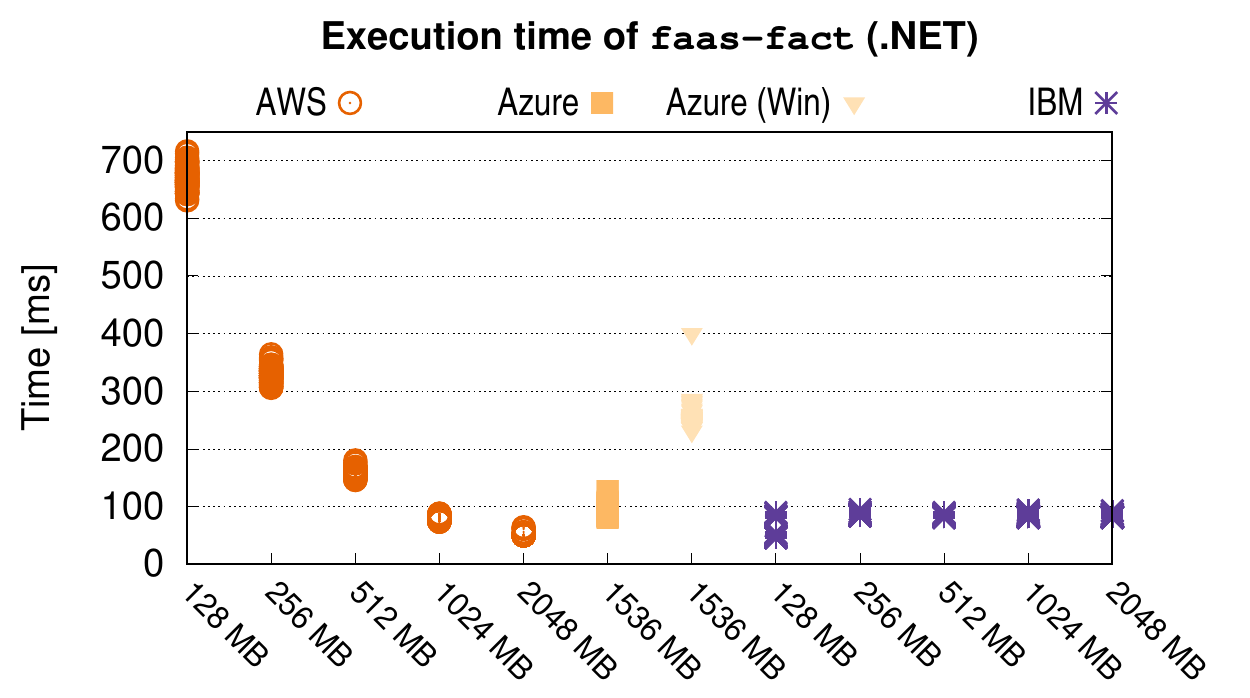}
\caption{Execution time of the \texttt{faas-fact} in .NET.}
\label{fig:general_dotnet_plot}
\end{figure}

When this happens and an invocation is scheduled for the just garbage-collected function, the cold start effect is clearly measurable with degrading effects on the measured latency.
%\vs{MAISSEN: can't understand this sentence, can you rephrase?} \mais{In order for the cloud provider to prepare the VM instance and deploy the user function code on it, some time can pass.}
\autoref{fig:azure_coldstart} shows the execution workflow for a cold start on Azure, as opposed to the warm case. 
Similar workflows (and effects on latency) exist on the other cloud providers.\footnote{Refer for instance to \url{https://cloud.google.com/functions/docs/bestpractices/tips} for information and best practices from Google.}
%\vs{MAISSEN: do you have the same results for the other providers? eg, why only Azure?} \mais{This is just exemplary done for Azure. Should be very similar on other clouds. see \url{https://cloud.google.com/functions/docs/bestpractices/tips} , little doc of providers because a 'disadvantage' of serverless}

Our results are averaged over ten executions for each configuration runtime/cloud. 
Functions are executed using 512\,\gls{MB} of memory and without external packages, to reduce the baseline footprint and the corresponding loading time to the bare minimum.
%The cold start latency has been tested ten times for each runtime on each cloud.
%Memory size was defined with 512 \gls{MB} and no packages were loaded into the function. 
We executed these tests on the following regions: \texttt{eu-central-1} for AWS, \texttt{west-europe} for Azure,  \texttt{europe-west1} for Google and \texttt{eu-de} for IBM. 
Note however that the chosen region has no impact on the cold start latency. 
We compute the cold start latency as the total request time minus the normal average latency.
On AWS and IBM, cold starts can be deteced with the help of instance IDs. 
On Azure and Google, online metrics of running instances were taken into account to verify cold starts. 
The challenge of cold start detection lead to a manual test execution and data manipulation.

\autoref{fig:coldstart_plot} summarizes our results using a box-and-whiskers plot.
It stands out that \gls{AWS} is overall the fastest, with an average cold start latency of only 335\,ms for Node.js, Python and Go. 
Results with .NET are worse, with latencies up to 1,739\,ms, likely due to the nature of the runtime and the compilation of C\#. 
On Azure, cold start latency is strictly more than 2 seconds and up to 5 seconds, with the exception of the combination .NET on Windows, which averages at 1,917\,ms. 
IBM compares similarly to \gls{AWS}.
Finally, Google Cloud consistently yields higher cold start latencies, in the 2-3 seconds range. 
\gls{IBM} exhibits a performs similarly to \gls{AWS} across the spectrum of supported languages, although the cold start latency is around 600\,ms higher for Node.js, Python and Go but similar for .NET.

%\vs{MAISSEN: what is this 'reclying' you refer to ? can you clarify?} \mais{Recycling meaning in this context deallocating a VM instance of the user and make it available to all users in the cloud's total pool}
On \gls{AWS} and \gls{IBM}, it took usually around 10 minutes of no activity for the computing instance to be recycled by the provider and re-inserted into the pool of available ones, and up to 20 minutes on Azure.
On Google, this time varied from 10 minutes up to 10 hours. %\mais{I said to you 12 hours as mentioned previously in this section. I meant also 10. I corrected this, you can remove this comment.}

\subsection{CPU-bound Benchmarks}
\label{sec:general_test}

%This test was performed to analyze the functions under normal circumstances, meaning that there is not much load respectively that the load can be handled well by the cloud provider. The objective is to compare the achieved performance of the clouds regarding execution speed and the thereby implicated costs. 
This benchmark evaluates how CPU-bound workloads behave across the different cloud providers.
% evaluation the \gls{CPU} factors test (see section \ref{sec:factors_test}) was executed. 
A function (\texttt{faas-fact}) is invoked every five seconds, using as parameter a large integer triggering a sufficiently challenging computation. 
We collect 100 results for every configuration (runtime, memory, cloud provider).
%The function was invoked every five seconds until a sample size of $n=100$ was reached. 
%This for each \gls{MB} configuration, each runtime and each cloud. 
%s input parameter $26'888'346'474'443$ was selected since with this number the function finished ahead of the timeout for low configurations and simultaneously was not too quickly finished for high configurations. 
We report these results for all cloud configurations and languages: Python (\autoref{fig:general_python_plot}), Node.js (\autoref{fig:general_node_plot}), Go (\autoref{fig:general_go_plot}) and .NET (\autoref{fig:general_dotnet_plot}).
%Figure~\ref{fig:general_python_plot} depicts our result for Python
%\vs{If we have time, we could include the same plots for the other languages. MAISSEN: could you give it a try, using the Python example as template to produce them? }

We note a few interesting facts.
On \gls{AWS} Lambda, the standard deviation is low, with a minimum value of 50\,ms using .NET and  2,048\,MB of memory, with very consistent execution times across all the different configurations. 
As expected, every doubling of the allocated memory results in halving the execution time, \ie, performance scales linearly with allocated memory~\cite{AWSLambdaConfig}. 

Azure's memory configuration of 1,536\,\gls{MB} suggests its performance to be in the 1,024-2,048\,\gls{MB} range.
Surprisingly, it is slower than 1,024\,\gls{MB} instances of the other clouds.  
Google Cloud Platform performs similarly to \gls{AWS}, although more scattered for 128\,\gls{MB} of memory. 
Google achieves better results than \gls{AWS} for several memory configurations (128, 256, 512 and 1,024\,\gls{MB}). % but not for 2048 \gls{MB}. For the first three times doubling the memory leads to halved execution time. 
Note however that the performance scaling behaves differently on the Google platform~\cite{GoogleFunctionsPricing}.
%But not entirely for the next two steps since Google does not scale \gls{CPU} completely linear to allocated memory \cite{GoogleFunctionsPricing}. 
The results we gathered on the \gls{IBM} platform suggests that memory allocation does not correlate with \gls{CPU} allocation in any way, with five different configurations performing very closely. 
This is remarkable since the pricing model of \gls{IBM} only accounts for the GB-seconds used. 
%\vs{MAISSEN: Is this true for all the languages? These resuls are only for Phython. Do you have the same tests for all the other languges? This means that a user could deploy his application always with the smallest memory option possible and would thereby get the same performance for the smaller price.} \mais{Yes this is true for all languages, see {https://github.com/Bschitter/benchmark-suite-serverless-computing/blob/master/results/3-general} files with name scatterplot\_general\_<RUNTIME>.png}
%The results and plots for this and the three other runtimes can be found on \href{https://github.com/Bschitter/benchmark-suite-serverless-computing/tree/master/results/3-general}{GitHub}

\begin{figure*}[!t]
\centering
\includegraphics[scale=0.7]{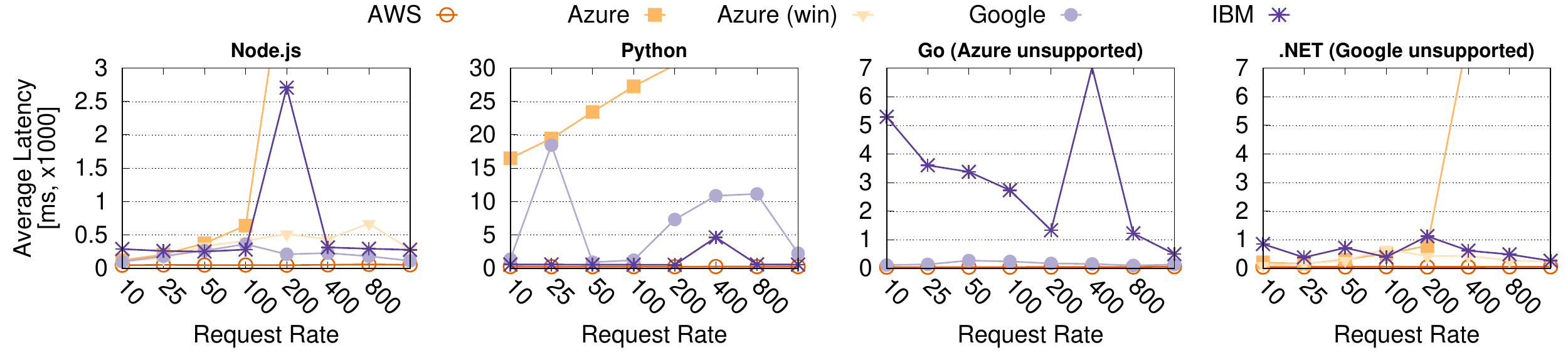}
\caption{Throughput/latency benchmark using \texttt{wrk2}~\cite{wrk2} and \sys's  \texttt{faas-matrix-mult} stress function.}
%\mais{Something's wrong. Did you intentially not plot Azure Windows for Node.js and .NET? Further, the .NET plot is wrong: AWS not plotted, Azure (Linux) has the color of AWS}}
\label{fig:load_test_latency_all}
\end{figure*}

\renewcommand{\arraystretch}{1.2}
\begin{table}[!t]
\setlength{\tabcolsep}{3pt}
\scriptsize
\centering
\caption{Latency: goal/target, results in the long-tail (below 90\% of the target requests/sec).}
\rowcolors{1}{gray!10}{gray!0}
\begin{tabular}{l|l|r|r|r} 
\rowcolor{gray!45}
\textbf{Cloud} & \textbf{Runtime} & \textbf{Req/s (goal)} & \textbf{Req/s (actual)} & \textbf{Req/s (\%)} \\ \hline
%\rowcolor{gray!25}
Azure & Node.js & 200 & 179 & 89.50\% \\ 
Azure & Node.js & 400 & 227 & 56.75\% \\ 
Azure & Node.js & 800 & 278 & 34.75\% \\ 
Azure & Node.js & 1000 & 327 & 32.70\% \\
Azure & Python & 10 & 6 & 60.00\% \\  
Azure & Python & 25 & 12 & 48.00\% \\ 
Azure & Python & 50 & 18 & 36.00\% \\ 
Azure & Python & 100 & 25 & 25.00\% \\ 
Azure & Python & 200 & 30 & 15.00\% \\ 
Azure & Python & 400 & 15 & 3.75\% \\ 
Azure & Python & 800 & 14 & 1.75\% \\ 
Azure & Python & 1000 & 15 & 1.50\% \\
Azure & .NET & 400 & 324  & 81.00\% \\
Azure & .NET & 800 & 407 & 50.88\% \\ 
Azure & .NET & 1000 & 512 & 51.20\% \\ 
Google & Python & 25 & 12 & 48.00\% \\ 
Google & Python & 200 & 168 & 84.00\% \\ 
Google & Python & 400 & 297 & 74.25\% \\ 
Google & Python & 800 & 575 & 71.88\% \\ 
IBM & Go & 25 &  21 & 84.00\% \\ \hline
\end{tabular}
\label{table:rps}
\vspace{-15pt}
\end{table}

\subsection{Throughput/Latency}
\label{sec:loadtest}

To understand the saturation point of the deployed services, we rely on \texttt{wrk2}~\cite{wrk2}, a constant throughput/exact latency HTTP-based benchmarking tool.
We configure this benchmark to issue function call invocations at increasingly high rates, from 10 up to 1,000 requests per second.
For each of the configurations, \texttt{wrk2} reports the average latency to handle the requests (\ie, the functions).
Between each configuration, the benchmark waits for sufficient time to process any request still possibly enqueued.
\autoref{fig:load_test_latency_all} shows our results for the \texttt{faas-matrix-mult} function and the 4 different programming languages under test.
%The load test is designed to benchmark the serverless functions up to 1000 requests per second. 
%This is carried out with the \gls{HTTP} benchmark tool \textit{wrk2} (see \cite{wrk2}). A function is first called with 10 requests per second for the duration of 1 minute. 
%Subsequently with 25, 50, 100, 200, 400, 800 and finally 1000 requests per second each time for one minute. 
%In between is a short break of 10 seconds to allow the function to process requests that are still queued. As test, the matrix function with a parameter of 100 is used. With this setup the matrix function should have an average execution time of about 100ms for Node.js, Go and .NET and around 250ms for Python. The graphic \ref{fig:load_test_latency_all} displays the average latency results grouped by runtime.
%The results are now discussed step by step. 

We observe that \gls{AWS} achieves stable response latencies for all the tested workloads. 
%This is very remarkable and a very good result and can be probably traced back to the low cold start latency and the good and consistent performance presented in section \ref{sec:general_test}.\\
Azure, on the other hand, shows more surprising results.
Whereas it can tolerate high loads on Windows OS, it performs very inconsistently on Linux.
The response latency grows almost linearly with the total requests per second injected and quickly saturates. 
That is a strong indication that none (or too few) new instances were allocated to handle the load. 
We investigated this behaviour further through the Azure portal Live Metrics Stream (not shown). 
When reaching 1,000 requests per second, only a total of 12 instances were deployed to serve .NET functions. % have been deployed (see figure \ref{fig:live_metrics_stream}). 
Additionally, only 500 requests per second were actually served, leaving a large percentage of requests in the waiting queue.
%\begin{remark}
%In the screenshot the request duration differs from the wrk2 results since that is most likely only considering execution time without queuing time.
%\end{remark}

%\begin{figure}[htp]
%\centering
%\includegraphics[scale=0.3]{bilder/Azure_Dotnet_1000.png}
%\caption[Azure Linux .NET Live Metrics Stream during 1000 RPS]{Azure Linux .NET Live Metrics Stream during 1000 RPS\\Source: screenshot Azure portal}
%\label{fig:live_metrics_stream}
%\end{figure}

\begin{figure}[!t]
\centering
\includegraphics[scale=0.7]{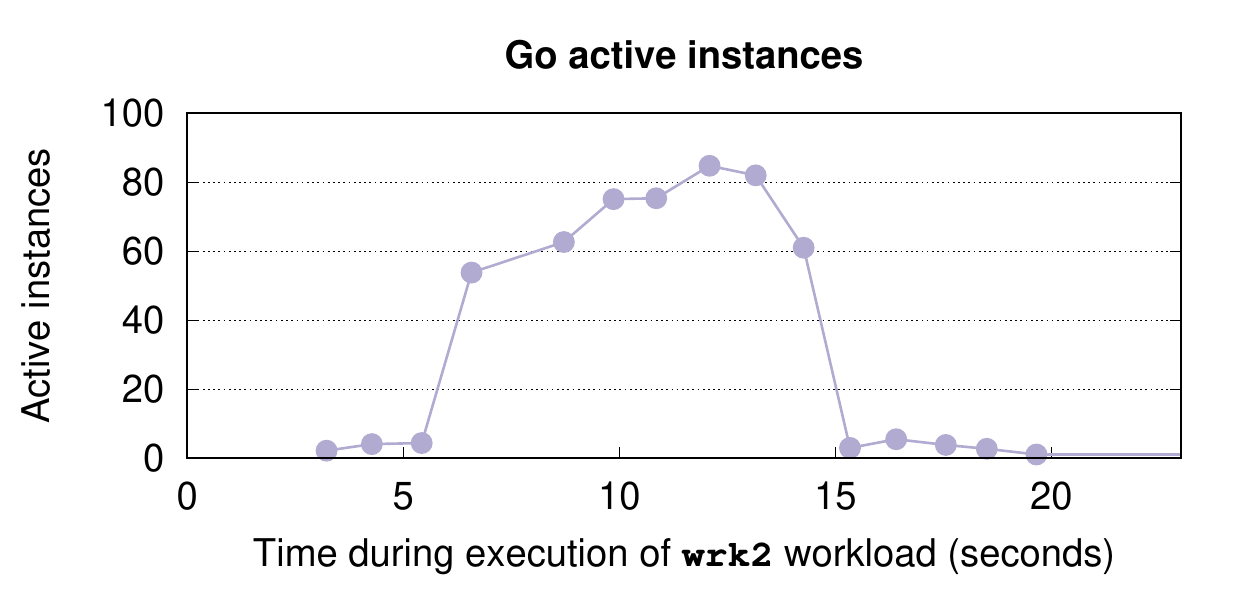}
\caption{Google Go: number of active compute instances during the wrk2 load test.}
\label{fig:google_graph_go}
\end{figure}

For Node.js and .NET, response latencies spike up to 24\,s and 18\,s, respectively. % and are not plotted in the graph for illustration purposes. 
Using Python, Azure managed to handle up to 200 requests per second, beyond which none of the requests could be served correctly.
Google can manage the load generally well, although latency increases slightly for Node.js and Go. 
However, Python functions deployed Google suffer from poor performance, in particular at higher request rates. 
Results suggest poor capacity in adapting quickly to flash-crowd requests, with the average request duration growing from 1,356 to 18,448\,ms when increasing the rate from 10 to 25 requests per second. 
%After that, this phenomena happens again but less drastically. 
%Since functions take on average a longer time to execute in Python, it seems and Google can scale longer running functions less good. %Overall the performance is good. 
We investigate further (\autoref{fig:google_graph_go}) by analyzing the scaling behaviour for instances running Google Functions.
The plot shows the number of instances allocated during the load test on Google, suggesting indeed that the auto-scalability features provided by the infrastructure might lead to poor results as requests start saturating the deployed instances.
%\vs{this was for Go}

%Finally, we discuss the results obtained using \gls{IBM} Cloud Functions.
%Lastly, the result of \gls{IBM} will be discussed. 
%Those outcomes differ from the others relatively much. 
On \gls{IBM} Cloud Functions, Node.js performs particularly poorly when compared to the other providers, while Python functions perform on par with \gls{AWS}.
Go functions achieve far worse than for other providers, contrasting with the cold start results that suggested similar behaviour. 
Further investigation is needed as part of future work to understand the reasons of this behaviour.
%On Node.js and .NET it can roughly keep up to the competition but operates a little slower and on Node.js it has a striking outlier. 
%The Python function on IBM nearly performs as well as on \gls{AWS}, but the same cannot be said about Go. 
%The Go runtime performed much slower in the load test as with the competitors, although as seen in the cold start test in section \ref{sec:coldstart} it has a similar cold start latency as Node.js and Python and as shown in the appendix table \ref{tab:general} it performs generally well on Go. 
%Thus this bad performance cannot be derived from this or other tests and it is possible that IBMs auto scaling mechanism does not work well for Go. 
%Of the amount of instances deployed, IBM delivers no statistics.
%\vs{MAISSEN: can you confirm I understood correctly the results in table 3?} \mais{Yes I think so, but bad wording for the last sentence. ->  In particular, we only report the cases which achieved 90\% and below of the target throughput (right-most column).}
\autoref{table:rps} reports for which cases the target throughput set by \texttt{wrk2} could not be reached within a 10\% margin, \ie, cases achieving at most 90\% of the target throughput (right-most column).

\begin{table}[!t]
    \caption{Pricing calculation: cost of the CPU-bound benchmarks computed by \sys according to the pricing models of each cloud. Execution times are rounded up to the next 100\,ms for AWS, Google and IBM~\cite{AWSPricing,GoogleFunctionsPricing, IBMPricing}\label{table:example2}}.
    \setlength{\tabcolsep}{3pt}
  \scriptsize
  \center
  \rowcolors{1}{gray!10}{gray!0}
  \begin{tabular}{r|r|r|r|r|r|r}
  \rowcolor{gray!45}
  \textbf{Memory} & \textbf{Exec. time}  & \textbf{Invocation} & \textbf{GB/sec} & \textbf{GHz/sec} & \textbf{BW} & \textbf{Total} \\
    \hline
  \rowcolor{gray!25}
    AWS  & & & & & &\\
    128\,MB & 8000\,ms &  2.00\$  &  166.67\$  &  --- & 3.43\$ & 172.10\$ \\
    256\,MB & 4000\,ms &  2.00\$  &  166.67\$  &  --- & 3.43\$ & 172.10\$ \\
    512\,MB & 2000\,ms &  2.00\$  &  166.67\$  &  --- & 3.43\$ & 172.10\$ \\
	1\,GB   & 1000\,ms &  2.00\$  &  166.67\$  &  --- & 3.43\$ & 172.10\$ \\
	2\,GB &  600\,ms &  2.00\$  &  200.00\$  &  --- & 3.43\$ & 205.43\$ \\
    \hline
  \rowcolor{gray!25}
    Azure  & & & & & &\\
    128\,MB & 1267\,ms &  2.00\$  &   25.34\$     &  ---        & 3.32\$ &  30.66\$ \\
    \hline
  \rowcolor{gray!25}
    Google  & & & & & &\\
    128\,MB & 7700\,ms &  4.00\$  &   24.06\$     &  154.00\$ & 4.58\$ & 186.64\$ \\
    256\,MB & 3200\,ms &  4.00\$  &   20.00\$     &  128.00\$ & 4.58\$ & 156.58\$ \\
    512\,MB & 1600\,ms &  4.00\$  &   20.00\$     &  128.00\$ & 4.58\$ & 156.58\$ \\
	1\,GB   &  900\,ms &  4.00\$  &   22.50\$     &  126.00\$ & 4.58\$ & 157.08\$ \\
	2\,GB &  800\,ms &  4.00\$  &   40.00\$     &  192.00\$ & 4.58\$ & 240.58\$ \\
    \hline
  \rowcolor{gray!25}
    IBM & & & & & &\\
    128\,MB & 700\,ms &  ---       &   14.88\$     &  ---        & ---      &  14.88\$\\
    256\,MB & 600\,ms &  ---       &   25.50\$     &  ---        & ---      &  25.50\$\\ 
    512\,MB & 600\,ms &  ---       &   51.00\$     &  ---        & ---      &  51.00\$\\ 
	1\,GB   & 600\,ms &  ---       &  102.00\$     &  ---        & ---      &  102.00\$\\
	2\,GB   & 700\,ms &  ---       &  238.00\$     &  ---        & ---      &  238.00\$\\
    \hline
  \end{tabular}
%\vspace{-10pt}
\end{table}

\subsection{Pricing}
\label{sec:pricing}
%\vs{check Exec. time column values}
We conclude our quantitative evaluation by showcasing one possible usage of \sys's billing calculator (\S\ref{ssec:billingcalc}).
To that end, we use it to evaluate the cost of the CPU-bound benchmarks if executed across all the serverless providers.

We assume a Python runtime and 10\,M function calls/months.
%Let's also assume 10 million function calls are invoked per month. \mais{Repetitve: 2x 10 m calls}
The allowed memory is set to a 100\,\gls{MB} function and a return payload of 4\,KB per call.
%With the fixed parameter of $26'888'346'474'443$ the function should not consume more than 100 \gls{MB} of memory. 
%Its return size is around 4 KB per call. 
%As execution times we take the average (rounded up to 100~ms if applicable) of the result.
We note that Azure supposedly charges the GB-seconds that the function actually used, rounded up to the next 128\,\gls{MB} step \cite{AzurePricing}. 
It seems therefore that the pricing model of Azure is similar to the one of \gls{IBM}, delivering the same performance independent of the memory size.
%Doing the equivalent calculations as in example 1, the following results are obtained as shown in table \ref{table:example2}.

\autoref{table:example2} shows the costs computed by \sys.
%\vs{MAISSEN: how much do they differ from the real costs - those actually charged? do we know?} \mais{No I don't know. Free tier and the low quantity did not allow to calculate that exactly. This is just an approximation.}
We observe that \gls{AWS} is particularly accurate: for every doubling of memory, execution time halves, keeping the costs constant.
Despite not being the cheapest option, it remains the most predictable one. 
%Azure is the cheapest provider: with th
%\gls{AWS} is an exemplar in showing how the pricing works. Since with each doubling of memory execution time halves the price is exactly the same because also the GB-Seconds remain the same. But with 170\$ to 200\$ it is not the cheapest in the list.\\
Azure is the cheapest cloud provider according to our computations.
Since this example function application only uses 100\,\gls{MB}, Azure also only charges for 128\,\gls{MB}: memory is dynamically allocated and only billed according to the next 128\,MB step.
%  \vs{MAISSEN: why only 1 row of results for Azure ? } \mais{Because memory is allocated and billed dynamically up to the needed next 128MB step. In this example the functions uses only 100 MB (because it is CPU bound) and Azure bills 128MB. Clear?}
%However the actual metrics of how much memory was used does not exist for Linux and is only depicted in a graph for Windows. But in the case this test is correct the cost-performance ratio is very good for low memory applications.\\
The costs of deploying Google Functions follows closely those of \gls{AWS}, especially for mid-range configurations.
For the least- and most-expensive configurations, AWS reveals to not be the best trade-off in terms of cost/performance.
Finally, IBM's billing method offers predictable costs, being roughly linearly dependant of the chosen memory configurations.
%Google generally conforms with the pricing results of \gls{AWS}. It positions itself in the same price region although it climbs in costs at 128 \gls{MB} and 2048 \gls{MB}. This can be traced back to the fact that it performed inconsistent and relatively seen slower for 128 MB, and for 2048 MB not much performance was gained. Therefore these two options are more expensive.\\
%IBM leans on Azure's method of calculating prices. It only charges for the defined and allocated GB-Seconds, but performs basically always the same. Hence, it gets approximately linearly more expensive with more memory allocation.
\subsection{Usability Considerations}

%In this section. the four serverless services will be compared. Two aspects were crucial for the evaluation: First, the previously discussed test and benchmark results and secondly, my personal opinion which was developed during this research and its implementation.
We finally report on some usability considerations regarding the cloud providers and their serverless offerings.

\textbf{AWS Lambda.}
Overall, this reveals to be the best-performing cloud provider. 
According to our continuous monitoring over several months, it provides consistent performance and reliability.
Management via the \gls{CLI} or Web-based portal is straightforward and efficient, even though setting up HTTP triggers remain convoluted.

\emph{Drawbacks:} 
\emph{(i)}~lack of official Docker image for the \gls{AWS} \gls{CLI}; 
\emph{(ii)}~deleting a function requires usage of \texttt{aws lambda list-functions}, which however only works on a per-region basis;
\emph{(iii)}~security counter-measures imposed by the AWS (\eg, changing public APIs more frequently than 30\,s) can reduce the productivity of the developers as well as the ability of quickly prototyping and testing functions.

\textbf{Azure Functions.}
Performance and usability are worse than AWS. 
The cold start latency of 2 to 4\,s makes it unfit for short-lived sessions. %, leading to increased costs \mais{cost should not be higher with cold starts. Clouds only bill when the function is actually running / executing code} especially under heavy workloads.
The Web-based portal provides useful insights on the real-time performance, as well as system insights from telemetry-based monitoring on the supporting computing instances.

\emph{Drawbacks:} 
\emph{(i)}~there is a discrepancy between the actions available via the official CLI and the Web-based portal;
\emph{(ii)}~additional tools are required also for local debugging (\eg, the Azure Functions core tools,\footnote{\url{https://github.com/Azure/azure-functions-core-tools}}), which seems counter-intuitive, and they are not shipped with any official Docker image;
\emph{(iii)}~the overall architecture, as well as the different generations of APIs, contribute in making the learning curve for Azure rather steep.

\textbf{Google Cloud Functions.}
This offering leverages a simple and clearly structured Web portal, as well as a consistent \gls{CLI} tool. 
Support for single-command operations is particularly helpful and practical. 
%Where the other clouds need multiple commands to get a function deployed and running it is just one simple command with Google (see figure \ref{fig:google_deploy}). The documentation is complete and clear.\\
While cold start latency is higher than AWS or IBM, it remains within usable limits. 
Regarding billing costs, Google is the least convenient operator.

\emph{Drawbacks:}
\emph{(i)}~the monitoring measurements available to the developers lag several minutes behind the real execution, preventing prompt interventions.
\emph{(ii)}~it supports a limited number of runtime systems, at the risk of being a less attractive deployment choice;
\emph{(iii)}~deployers are left with very few configuration options, preventing (by design) fine-tuning operations by the clients.
%\vs{MAISSEN: anything else we could mention here?} \mais{maybe that only 3 runtimes are supported, pricing calculated more granularly which makes it harder for a user to estimate / understand, not that many configuration / control over the service (but this can be seen as good or bad)}
%In regards to performance Google functions is good. The cold start latency is relatively high compared to \gls{AWS} and \gls{IBM}. 
%Execution times are similar as on \gls{AWS} and scaling during the load test was okay. It seems to work well for fast executing functions but not so good for slower functions (higher than 0.5 seconds) which could be a deal breaker. On paper Google is the most expensive of them all.\\
%In the Google portal there are some nice graphs to monitor number of allocated instances, execution time, number of invocations and memory used per call. It is not properly real time as the one of Azure since it lags behind a few minutes.

\textbf{IBM Cloud Functions.}
The current public release of IBM lacks straightforward and well-structured documentation.
The cold-start performance suggests this provider to be a good candidate for quick tests, where an application can start invoking functions with short waiting times.
%According to our experimental evaluation, IBM achieves the lowest memory footprint \mais{Why? I don't understand.}.

\emph{Drawbacks:}
\emph{(i)}~the official IBM \gls{CLI} can only be used if there is a corresponding cloud foundry support for the intended region (currently the case for all public regions except for Tokyo);
\emph{(ii)}~the support for the Go runtime is rather poor, resulting in the worst performance for this configuration;
\emph{(iii)}~some (important) features are left undocumented, such as the limit of 3,000 request per minute authorized for normal accounts.

\section{Lessons Learned}
\label{sec:lessons}

This study has shown that serverless computing in the form of \gls{FaaS} can deliver good performance at a reasonable price.
It is fairly easy to set up, deploy and execute code in the cloud for the end user, as compared to \gls{VM}s or even physical servers.
With serverless computing, companies can focus on building their business logic instead of maintaining operating systems and servers on premises, which requires expertise and is not always affordable for small companies.
One of the most important features of \gls{FaaS} is is auto-scaling capability, which take away from the deployer the burden of dimensioning the system and allocating resources.

Through the implementation and evaluation of \sys, we notably encountered the following limitations of the FaaS paradigm. 

First, one is heavily limited by the offering of specific cloud providers, including available runtime systems and programming languages, supported triggers, third party  services integration, quotas or maximum memory. 
Users cannot easily overcome such limitations as the full application stack is managed by the provider.

Second, providers can force an application to migrate from a specific runtime version to an arbitrarily different one (\eg, when upgrading their system from Node.js v6 to v8).
Despite backward compatibility, functions might break, forcing developers to keep up with the pace imposed by the cloud provider.

Third, developers face the risk of vendor lock-in by depending on proprietary features.
This might become a serious issue when considering application upgrades and potential provider migrations.

Finally, the lack of deployment standards makes the FaaS paradigm still a largely experimental playground.
Providers offer custom APIs (\eg, request/response objects), which inevitably vary across cloud offerings.
A common framework supported by a standard would certainly improve adoption and benefit the whole ecosystem.

\section{Conclusion and Future Work}
\label{sec:conclusion}

Motivated by the lack of common FaaS framework or standard platform to evaluate serverless platforms, this paper presented \sys, a user-space application that facilitates performance testing across a variety of serverless platform.
\sys is an open-source benchmark suite that is easy to deploy and provides meaningful insights across a variety of metrics and serverless providers. 

We envision to extend \sys along the following directions.
First, we will expand the set of supported languages and runtime systems, also by integrating contributions from the open-source community.
Second, we plan to maintain a continuous, online deployment of the \sys, publicly accessible, in order to build a larger dataset of historical measurements that will be released to the research community.
We finally intend to include native support for Docker images to ship functions, in order to facilitate integration with services such as Google Cloud Run~\cite{cloudrun}.

%Although this benchmark suite covers the most important aspects it can be developed and be improved further. Apart from general improvements and optimizations more runtimes respectively languages and cloud providers could be supported, more different tests could be implemented and provided. 
%In addition, a plotting integration (e.g. with R) would be useful as Grafana does not provide much exporting and plotting capabilities and is rather focused on live monitoring.\\
%There could be a test regarding continuous deployment and inspect the behaviour of the platform. Besides, a special focus could lay on deploying and testing Docker images because that might be an important aspect and trend in the future, considering there are no runtime restrictions of the provider. Several providers already offer to deploy just a Docker image (AWS, Azure, IBM) and Google has its service Cloud Run which has become generally available on November 14, 2019 \cite{cloudrun} and is a fully managed, serverless container platform.\\ 
%Moreover, the load test should probably be carried out with different function execution times to get a better understanding of the possible connection between scaling and execution time, if there is any at all. 
%Also, a real world application test would be beneficial and demonstrate the usability of serverless computing.

%\newpage
\vspace{-8pt}
\bibliographystyle{ACM-Reference-Format}
\bibliography{bib}
\end{document}